\documentclass[12pt]{article}

\usepackage{amsmath}
\usepackage{graphicx}
\input epsf



\makeatletter

\renewcommand\section{\@startsection{section}{1}{\z@}
                                   {-3.5ex \@plus -1ex \@minus -.2ex}
                                   {2.3ex \@plus .2ex}
                                   {\normalfont\large\bfseries}}
\renewcommand\subsection{\@startsection{subsection}{2}{\z@} 
                                   {-3.25ex\@plus -1ex \@minus -.2ex}
                                   {1.5ex \@plus .2ex}
                                   {\normalfont\normalsize\bfseries}}
\renewcommand\subsubsection{\@startsection{subsubsection}{3}{\z@}
                                   {-3.25ex\@plus -1ex \@minus -.2ex}
                                   {1.5ex \@plus .2ex}
                                   {\normalfont\normalsize\bfseries}}
\renewcommand\paragraph{\@startsection{paragraph}{4}{\z@}
                                   {3.25ex \@plus1ex \@minus.2ex}
                                   {-1em}
                                   {\normalfont\normalsize\bfseries}}

\makeatother

\newcommand{\Tr}{{\rm Tr}\,}
\renewcommand{\Im}{{\rm Im}\,}
\newcommand{\be}{\begin{equation}}
\newcommand{\ee}{\end{equation}}
\newcommand{\ba}{\begin{eqnarray}}
\newcommand{\ea}{\end{eqnarray}}
\newcommand{\nl}{\nonumber \\}

\newcommand{\DD}{{\cal D}}

\newcommand{\qs}{\,/\!\!\!\!q}
\newcommand{\ps}{\,/\!\!\!p}
\newcommand{\es}{\,/\!\!\!\epsilon}
\newcommand{\pslash}{\,/\!\!\!p}
\newcommand{\qslash}{\,/\!\!\!q}
\newcommand{\sgn}{{\mbox{sgn}}}
\newcommand{\nr}[1]{(\ref{#1})}
\newcommand{\OO}{{\cal O}}

\def\MSbar{\overline{\textrm{MS}}}

\renewcommand{\Im}{{\rm Im}\,}

\newcommand{\PiR}{\Pi_\textrm{R}}
\newcommand{\aslash}[1]{#1\!\!\!\!\slash}

\def\AA{\mathcal{A}}
\def\GR{G_R}
\def\GA{G_A}

\def\ge{\gamma_\textrm{E}}
\def\b#1{\langle{#1}\rangle}
\def\l{\langle}
\def\r{\rangle}

\begin{document}
%\begin{flushright} {IAS 2009}\end{flushright}

\begin{center}
\vspace*{5mm}
{\Large\sf
{  Loops and trees
}
}

\vspace*{5mm}
{\large S. Caron-Huot}

\vspace*{5mm}
       Institute for Advanced Study, Princeton NJ 08540, USA
\vspace*{5mm}

{\tt schuot@ias.edu}

\vspace*{9mm}
{\bf Abstract} 
\end{center}
\vspace*{1mm}
\noindent 
We investigate relations between loop and tree amplitudes in quantum field theory
that involve putting on-shell some loop propagators.
This generalizes the so-called Feynman tree theorems satisfied at 1-loop.
Exploiting retarded boundary conditions, we give a generalization to $\ell$-loop expressing
the loops as integrals over the on-shell phase space of exactly $\ell$ particles.
We argue that the corresponding integrand for $\ell>2$ does not involve
the forward limit of any physical tree amplitude,
except in planar gauge theories. In that case we explicitly construct the relevant physical
amplitude.
Beyond the planar limit, abandoning direct integral representations, we propose
that loops continue to be determined implicitly by the forward limit of physical connected trees,
and we formulate a precise conjecture along this line.
Finally, we set up technology to compute forward amplitudes in supersymmetric theories,
in which specific simplifications occur.
\setcounter{tocdepth}{1}
\tableofcontents

\setcounter{equation}{0}
\section{Introduction}

Relations between loop and tree amplitudes are taking an increasingly important place in
modern quantum field theory.
To find such relations is actually a fairly old program \cite{eden}.
Modern interest in it stems from the striking simplicity of on-shell tree amplitudes in gauge theories
compared to the corresponding Feynman diagrams, which one aims to exploit in loop computations.
The unitarity-based methods,
pioneered in the mid-nineties in work by Bern, Dixon, Dunbar and Kosower \cite{bddk},
may be said to successfully meet this goal.
These methods have now been successfully applied to multi-parton loop amplitudes
in supersymmetric theories, but, also, in QCD.
A non-exhaustive sample of references is \cite{cachazo,anastasiou,forde,bern07,berger}.

The unitarity-based methods, generally speaking, exploit the fact that the
discontinuities of a loop amplitude, or equivalently its unitarity cuts in all possible channels,
suffice to determine it.
The accuracy to which these cuts must be known may ultimately depend upon the theory;
for instance, in QCD it is necessary to know the cuts in a neighborhood
of $D=4-2\epsilon$ dimensions \cite{Badger}, while in supersymmetric theories,
4-dimensional data suffice \cite{bddk}.
Independent of these considerations, what is striking about these methods is that
to find what a loop integrand was, one has to integrate it across unitarity cuts.
In this paper we investigate possible more direct, purely algebraic,
ways to determine the loop integrands.

The idea is to try to determine loop integrands from their poles.
Such poles are caused by loop propagators becoming on-shell,
and so have the interpretation of on-shell particles circulating inside the loop.
The corresponding data thus consists of on-shell tree scattering
amplitudes in the forward limit, that is, zero momentum transfer for the particle in the loop.
We will try to shed light on the physical origin of such relations,
by trying to directly express loop amplitudes as phase space integrals over this data.
Instead of unitarity, the relevant physical principle proves to be causality.

The 1-loop version of this story was, in fact, worked out long ago by Feynman,
and so it is sometimes known as ``Feynman's tree theorem'' \cite{Feynman}.
It is useful to review it here.
The first step is to use essentially
contour integration in the frequency plane, to localize the
frequency integrals to the poles of propagators.  This succeeds in reducing any loop to
on-shell phase-space integrals.
The second step is to organize the combinatorics of the remaining part of the graphs,
which are trees.
Feynman could show at 1-loop that precisely the forward limit of physical tree amplitudes entered.%
\footnote{Feynman used this result to exhibit the failure
 of longitudinal gauge modes to decouple, in gravity and Yang-Mills loops.
 To cancel them at 1-loop in an off-shell formulation
 he introduced particles now known as Faddeev-Popov ghosts.
 Somewhat amusingly, we are following the exact same road backward:
 we aim to move toward an on-shell formulation,
 where amplitudes are simpler, and, in particular, ghosts never appear.
}

The contour integration step is actually subtle:
it is not very naturally compatible with the familiar $i\epsilon$ prescriptions.
To see this, it is useful to follow Feynman and replace contour integration
in the frequency plane by the one identity%
\footnote{We ignore any issue with convergence
 and arcs at infinity, which are absent in dimensional regularization.}
\be
 \int d^4q\, \GR(q)\GR(q+p_1)\cdots \GR(q+p_n)~q^{\mu_1}\cdots q^{\mu_j} = 0,
 \label{grloop}
\ee
in which the $\GR$ are the retarded propagators:
\be
 G_R(p) = \frac{-i}{-p_0^2 +\vec{p}^2  +m^2 -i\epsilon p^0}. \nonumber
\ee
In words, ``a closed loop of retarded propagators is zero.'' This should be rather obvious
in the time domain.  At the same time,
in frequency space, this follows from contour integration since
the poles of all retarded propagators lie below the real $q^0$ axis allowing
the $q^0$ contour to be deformed to infinity.
This identity can be used to compute a loop with time-ordered propagators $G_T$, by
substituting
\be
 \GR=G_T+G^<  \nonumber
\ee
into it, where $G^<$
is an on-shell propagator accounting for the mismatch in the $i\epsilon$ prescriptions
(for precision, the ``$<$'' superscript refers to its support being
 on negative frequencies).
Schematically, for a loop containing $m$ propagators this leads to
\be
 \int d^Dq ~(G_T)^m = \int d^Dq \left[ (G_T+G^<)^m - (G_T)^m \right].  %\label{nrbasic}
 \nonumber
\ee
Expanding out the bracket gives a series of terms each one of which contains at least one
on-shell propagator.
After summing over Feynman diagrams, this can be organized in terms of physical data.

We will not use this form of the tree theorem.
This form contains $(2^m-1)$ terms on the right-hand side
(not all of which may be kinematically allowed, of course), which can be a lot.
The bulk of them comes from just changing $G_R$'s to $G_T$'s and
can be interpreted as just changing boundary conditions.
Rather, we will avoid such contributions by considering observables
with retarded boundary conditions.
One can then hope that a simple analytic continuation will relate these observables to
time-ordered ones, and this will indeed be the case.
In any case, since the identity at play is due to causality,
it makes perfect sense to try it on causal observables.

What should we use as ``causal observables''?
Since retarded boundary conditions are a preferred choice in classical field theory,
we turn to there for inspiration.
In that context, a natural question is, under the influence of small perturbation,
how do fields change in the future?
Response functions answer this question, if we define an $n$ point response function
to be the solution to the field equations with retarded boundary conditions and $(n-1)$ infinitesimal sources,
the $n$-th leg becoming the observed field.
Perturbatively, classical response functions are computed just by summing over all tree diagrams as usual,
but with the replacement $G_T\to G_R$ for every propagator.  The time flow in every retarded propagator,
evidently, is such as to point toward the observed field.
In configuration space a response function vanishes unless
all $(n{-}1)$ sources are in the past lightcone of the $n$-th leg.
Since the same question can be asked in a quantum theory, response functions can also be defined
at the quantum level.
Response functions may be familiar to some reader from many-body or finite-temperature contexts \cite{Chou};
for the benefit of other readers, their definition and key properties will be reviewed below.
For the moment, the most important aspect to keep in mind is their space-time support.

For a 1-loop response function the tree theorem takes the particularly concise form:
\be
 R^\textrm{1-loop}_n(p_1,\ldots,p_n) =
  \sum_h (-)^{2h}
 \int_q \pi\delta^{+}(q^2{+}m^2_h)~ R^\textrm{tree}_{n+2}(p_1,\ldots,p_n;q^{h},-q^{-h}),
 \label{1loop}
\ee
where
\be
 \int_{q} \equiv \int \mu^{2\epsilon}\frac{d^Dq}{(2\pi)^D} %\pi\delta(q_1^2-m_1^2) \theta(q_1^0)
 \nonumber
\ee
is the ordinary loop integral
in $D=4{-}2\epsilon$ dimensions and the sum is over the helicities $h$ (or spin) of the particles.
This form was obtained relatively recently by Catani et al. \cite{catani} using contour integration arguments.
Morally, the integrand is to be evaluated in $D$ dimensions,
if one is to work within a consistently regularized theory.
Both sides of the equation involve retarded amplitudes, for which there is one special
``future-directed'' leg (the same on both sides).
We do not indicate this leg explicitly and the semi-colon refers to the fact that the
last two legs on the right are amputated.
For a loop diagram with $m$ propagators, this form contains exactly $m$ terms.

It is actually easy to understand why this form contains only $m$ terms.
This is due to the boundary conditions.
On-shell propagators have no notion of time flow so a product of
two subgraphs separated by on-shell propagators could not possibly obey retarded boundary conditions.
There must always be a connected tree of retarded propagators.
This argument is very powerful, and it applies to any loop order:
at $\ell$-loop, at most $\ell$ propagators can get cut in a causality-preserving fashion.
On the other hand, using contour integration in $\ell$ frequency variables,
one can always cut at least $\ell$ propagators.
The conclusion is that, any $\ell$-loop amplitude can be represented
as a phase space integral over exactly $\ell$ on-shell particles, times connected trees!
Investigating such formulas is the point of this paper.

At 1-loop \nr{1loop} gives rise to a quite elegant picture of the quantum vacuum,
as being filled with on-shell particles, whose interactions with us give rise to quantum corrections.
We cannot see them, because they uniformly fill space-time,
like a bosonic version of the ``Dirac sea.''  But although invisible in that sense, they
do generate tangible quantum effects.
In this paper we will see quite explicitly that such a picture fails at 2-loop in general,
a physical interpretation of the connected trees being unavailable.
An exception will be for planar gauge theories, where introducing a very specific set of
color-correlations for the vacuum particles will restore this interpretation.
This nicely fits within the idea that planar theories are ``classical'', albeit in an unusual sense.

This paper is organized as follows.
In section 2 we attempt to ``guess'' the higher-loop extension of \nr{1loop}
based on causality and unitarity.
A clash between those two principles will be argued to obstruct physical integral representations.
Specifically, we will find a unique integral representation
but it will not involve a physical combination of tree diagrams.
This explains why a 2-loop extension of \nr{1loop} has never appeared in the literature.
However, we will see how, in the planar limit, the unphysical sum
over trees becomes equivalent to a physical one.
In this paper we also make a proposal regarding non-planar theories.
Abandoning Minkowski-space phase-space integrals in theses case, we propose
that the forward limits of physical connected trees
continue to implicitly determine loop integrals.
For the reader's convenience, a precise statement in the form of a conjecture is given at the
end of section 2.

Section 3 contains a pedagogical introduction to response functions
and causal boundary conditions in quantum field theory.  Their relation
to time-ordered functions is also explained.
An efficient formalism to compute them, the Schwinger-Keldysh formalism, is introduced.
Quite pleasingly, we will see that the Feynman rules of this formalism,
when expressed in the appropriate field basis,
directly generate the 1-loop formula \nr{1loop} without requiring any work.

In section \ref{sec:newres} we apply this formalism to 2- and 3-loop diagrams.
A certain special vertex entering the Schwinger-Keldysh formalism will enter,
which will turn out to be related to the obstruction found in section 2.
The evaluation of these diagrams will establish
the 2-loop formula deduced there involving unphysical trees.
Starting from this formula, we will prove the physical one valid in planar gauge theories.
This section concludes with formal manipulations supporting our ``uniqueness conjecture'',
which exploit complexified Minkowski space.

In section \ref{sec:ex} we work out a simple 2-loop example of the formalism:
the 2-loop vacuum polarization in QED (or QCD).
We calculate it starting from physical
on-shell tree amplitudes.

Finally, some technology to efficiently compute the forward amplitudes in supersymmetric
theories is discussed in section \ref{sec:susy}.  In this case the definition of
the forward limit of amplitudes for S-matrix elements, as opposed to correlation functions,
is fully worked out.  This is simplified by supersymmetry cancellations.
Another simplification is the fast decay of amplitudes at large loop momenta,
which allows the use of factorization or recursion formulas.
A few concrete examples are given in the simplest case of planar $\mathcal{N}=4$
supersymmetric Yang-Mills, which is clearly singled out as the simplest theory to
which to apply this formalism.

The tree theorems, as well as nontrivial implications of causality at tree-level,
have appeared in the recent works \cite{catani,brandhuber,vaman}.

%Some recent works are closely related to ours.
%The idea that a loop integral is completely determined by its poles
%is the basis of the so-called OPP-method \cite{OPP}, which
%allows (so far) for the computation of 1-loop amplitudes in terms
%of tree amplitudes.

\section{Causality and unitarity}
\label{sec:guesswork}

We have just argued that causality implies the existence of on-shell integral
representations of loop amplitudes.
We now wish to argue that by combining this information with unitarity,
the form of these relations is uniquely determined.

At 1-loop, \nr{1loop} is pretty much the only answer one could get:  one has to open
exactly one propagator in the loop.
For a 1-loop $n$-point amplitudes, this leaves out a ($n+2$)-point tree amplitude as in \nr{1loop}.
Assuming that combinatorics work out nicely, this tree amplitude has better be the forward limit
of a physical tree amplitude.
We will check below that this is the only possibility
consistent with unitarity, but let us move on to 2-loop.

%(A brief word on our notations.
% The semi-colon in \nr{1loop} means that the last two legs are amputated.
% So far we have not specified whether
% the other legs are amputated or not, and all formulas apply to either case.
% We do not indicate which leg is the retarded one.)

A naive 2-loop exponentiation of the pattern would yield
\ba
 R^\textrm{2-loop, classical}_n(p_1,\ldots,p_n) &=& 
  \frac{1}{2!}
 \prod_{i=1}^2 \left[\sum_{h_i} (-)^{2h_i}
\int_{q_i} \pi\delta^+(q_i^2{+}m_{h_i}^2)\right]
\nl
&\times&
   R^\textrm{tree}_{n+4}(p_1,\ldots,p_n;q_1^{h_1},-q_1^{-h_1},q_2^{h_2},-q_2^{-h_2}).
 \label{NAIVE}
%\nonumber
\ea
This turns out not to be correct, meaning that we have to think harder.

The reason this is not correct is the following.
This formula amounts to treating the vacuum as a collection of classical fluctuations:
 in it the fields obey the classical equations of motion, but they
 are averaged over a distribution of initial conditions.
 The distribution that exponentiates to \nr{NAIVE} is a Gaussian,
and there can be nothing wrong with that if the initial conditions are set at some early time
where the interactions are adiabatically shut off.
Furthermore, the two-point function read off from \nr{1loop},
$(-)^{2h}\pi\delta^+(q^2)$,
gives each oscillator
mode an average energy $(-)^{2h}\frac12\hbar\omega$ which would seem to be
one's best guess at a classical description.
The problem, now, is that \nr{NAIVE} does not address
any of the well-known problems of the classical theory.
For instance,
the energy distribution $\hbar \frac \omega2$ is not
dynamically stable -- it would attempt to thermalize -- making this expansion
not even self-consistent.
A second profound problem is energy positivity: in a classical theory endowed with
``vacuum'' fluctuations, energy would not remain in the ``vacuum'' but would
transfer itself to any particle present on top of it.
This, of course, does not happen for zero-point energy in quantum field theory,
it being impossible to extract energy from a state of lowest energy,
in agreement with observations.

\begin{figure}
\begin{center}
\includegraphics[width=10cm]{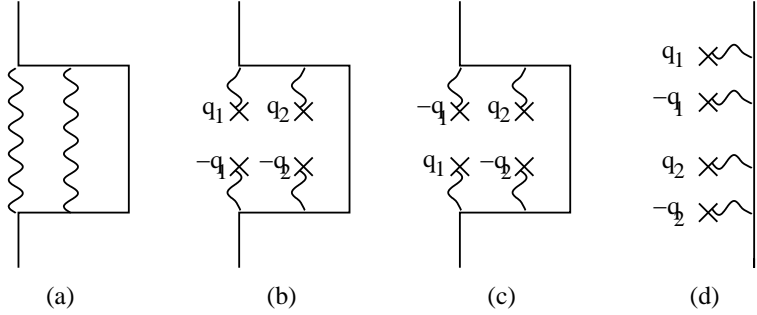}
\end{center}
\caption{
 (a) A 2-loop electron self-energy diagram.  The photon energies flowing in it can be
 parallel or anti-parallel as in (b) and (c), each case having to be considered separately in order
 to respect energy positivity.
 In some other diagrams (d), the energies
 in two cut propagators cannot be compared as they do not add nor subtract in any denominator.
}
\label{fig:electron}
\end{figure}

Since energy positivity is closely related to unitarity (it being the familiar
statement that energies along a unitarity cut are positive), this second problem implies
that \nr{NAIVE} violates unitarity in a very bad way.
Let us expound this in more detail.
Consider the example of an electron's self-energy in QED.
According to the classical model \nr{NAIVE},
the electron would acquire a large collisional imaginary part
through inverse Compton scattering against classical fluctuations of the electromagnetic background.
Diagrammatically this is the 2-loop graph shown in Fig.~\ref{fig:electron} (a).
Note that this is 2-loop graph, not contradicting the validity of a classical picture at 1-loop.
Taking the imaginary part
of this self-energy (which for retarded boundary conditions
 is also the lifetime of the particle)
 gives, for the term in \nr{NAIVE} corresponding to (b) and (c),
  graphs in which the electron propagator is cut so this gives a unitarity cut.
Now if both (b) and (c) are included, as dictated by \nr{NAIVE},
there will be a contribution in which the two photon energies have opposite signs
in the loop, corresponding to an absorption-emission process,
that is, inverse Compton scattering.
This produces the linearly divergent decay rate
 \footnote{
 The complete evaluation of \nr{NAIVE} for Fig.~\ref{fig:electron} (a)
 also includes the vacuum fluctuations of the electron's field.
 This modifies the coefficient but does not remove the linear divergence.
}
\be
 \Gamma \sim e^4\Lambda_\textrm{UV}.  \nonumber
\ee
Energy from the classical ``vacuum'' is transferred to the electron at this rate.
Evidently, such a huge scattering rate would
destroy any atomic system almost instantly;  \nr{NAIVE} fails at even the most
basic qualitative level.

Ordinarily, Feynman's $i\epsilon$ prescription guarantees the absence of such unphysical contributions,
which is why one never worries about them.
What happens here is that, to cut-open the loop, we had to use causality which clashes
with energy positivity (and with Feynman's $i\epsilon$ prescription).
The price to pay for
working with causality-friendly propagators is that the
unphysical region no longer cancels in individual diagrams.
It must, of course, cancel in the end through some other mechanism.
After all, the underlying quantum field theory is the same and we are merely asking a different sort
of question.

The only way we can correct this,
considering Minkowski-space integral representations,
is to correlate the signs of the energies $q_1^0$ and $q_2^0$.
This forces us to use a nonstandard weighting of tree diagrams
so as to avoid the automatic symmetry $(q_i\leftrightarrow -q_i)$ for each of the $q_i$.
The most general weighting can be described as follows.
For any tree diagram with four special external legs, $(q_1,-q_1,q_2,-q_2)$,
there are two unique oriented paths which link $q_1$ to $-q_1$
and $q_2$ to $-q_2$ (see Fig.~\ref{fig:electron}).
If they share a propagator with the same orientation (that is
if some propagator carries momentum $(q_1{+}q_2{+}p)$ for some $p$),
we say that $q_1$ and $q_2$ are parallel, and if there is minus sign,
we say that they are anti-parallel.
It may also happen that they do not share any propagator, in which case we
say that $q_1$ and $q_2$ to be non-comparable.
One linear combination of these cases must respect unitarity.
Let us state it, and then explain it:
\ba
 R^\textrm{2-loop}_n(p_1,\ldots,p_n) &=& 
  \frac{1}{2!}
\prod_{i=1}^2\left[ \sum_{h_i} (-)^{2h_i}
\int_{q_i} \pi\delta^+(q_i^2{+}m_{h_i}^2) \right]
\nl
&\times& \sum_\textrm{connected trees}
 R_{n+4}(p_1,\ldots,p_n;q_1^{h_1},-q_1^{-h_1},q_2^{h_2},-q_2^{-h_2})
\nl &
\times&
 \left\{\begin{array}{ll}
 \displaystyle  1+1/3,& \mbox{$q_1$ and $q_2$ parallel}, \\
 \displaystyle  1-1/3,& \mbox{$q_1$ and $q_2$ anti-parallel}, \\
 \displaystyle  1,& \mbox{$q_1$ and $q_2$ non-comparable}.
 \end{array} \right. \label{R2loop}
\ea

The curious factors $\frac13$ ensure unitarity.
To see this, consider a three-particle unitarity cut.  Essentially, $\frac13$
is because only two energies in the unitarity cut can be controlled in a given term.
Let us focus on the unphysical region in which two propagators have positive energy
and the third has negative energy.
In this region, for one term in \nr{R2loop}, $q_1$ and $q_2$ will have positive energy
making it a ``parallel'' configuration.
In two other terms, $q_1$ and $q_2$ will have opposite energies making them
``anti-parallel''.  Accounting for the sign flip for the imaginary part of a retarded propagator
at positive and negative energies, the total imaginary part from this region will be
\be
  -(\mbox{parallel)} + 2(\mbox{anti-parallel}). \label{energypos}
\ee
Unitarity (or energy positivity) requires that this be zero, which fixes
the ratio of the first two terms in \nr{R2loop}.
The unitarity cuts in the all-energies-positive region
 (which is the same for a retarded and a time-ordered function, in the absence of a fourth propagator)
determine the remaining coefficients; note that the average of the coefficients just matches
the classical expectation.
The conclusion is that \nr{R2loop} is the only Minkowski-signature
integral representation consistent with causality and unitarity at 2-loop.

Related problems with a classical model at 2-loop were also found by \cite{holdom}.

At 1-loop, the same computation gives a ``1'' instead of a ``2'' in \nr{energypos},
showing that \nr{1loop} duly respects unitarity.

Formula \nr{R2loop} will be proved in section \ref{sec:newres} using a method not relying on unitarity.
But this is not yet the formula we are looking for:
its integrand is unphysical --
we find that longitudinal polarizations and ghosts do not even cancel in it.\footnote{
 They do, of course, cancel when the integral is taken, for gauge-invariant observables,
 as usual.}

\subsection{Planar gauge theories}

The situation proves to be much better in planar gauge theories.
This is for purely kinematics reasons:
the color flow in a planar tree diagram is so constraining that, fixed the external momenta
and color structure, the momentum flow in all channels is completely fixed.
Hence one can replace the condition on the momentum flow in \nr{R2loop}
by one on the external color structure.
Equivalently, the problem with the instability of the classical vacuum is solved
by a universal distribution of color correlations,
one which can presumably be computed loop order per loop order.

Specifically, consider a gauge theory with a double-line notation,
such as $U(N)$ gauge theory, and consider the large $N$ limit
with all fields in the adjoint representation.
Tree amplitudes can be (uniquely) expanded as
\be
 R_n(1,\ldots,n) = \sum_{\sigma\in S_{n-1}}
 2g^{n-2} \Tr\left[t^{a_1} t^{a_{\sigma(2)}}\ldots t^{a_{\sigma(n)}}\right]
 \hat{R}_n(1,\sigma(2),\ldots,\sigma(n)).
\ee

Extra legs added to compute planar loops are color-adjacent.
At 1-loop, for instance, they can be inserted in $n$ ways,
$\hat{R}_{n+2}(\ldots,q_i,q_1,-q_1,q_{i+1},\ldots)$.
At 2-loop there are exactly $n(n+1)/2$ possibilities.
They fall into two distinct classes:
two disjoint pairs
$\hat{R}_{n+4}(\ldots,q_1,-q_1,\ldots,q_2,-q_2,\ldots)$
or two nested pairs $\hat{R}_{n+4}(\ldots,q_1,q_2,-q_2,-q_1)$.
It is easy to see that the first case includes all Feynman diagrams where
$q_1$ and $q_2$ are parallel,
while the nested case includes all
diagrams in which they are anti-parallel
(the two cases get exchanged, evidently, when the sign of $q_1$ or $q_2$ get exchanged).
Hence, following the above considerations, an Ansatz
can be constructed consistent with both causality and unitarity:
\ba
 \hat{R}_n^\textrm{2-loop,planar}(p_1,\ldots,p_n) &=&
  N_c^2
\prod_{i=1}^2 \left[ \sum_{h_i} (-)^{2h_i}
\int_{q_i}
\pi\delta^+(q_i^2{+}m_{h_i}^2) \right]
\nl &\times& \left[
 \sum_{i\leq j}
 \hat{R}_{n+4}(\ldots,p_i,q_1,-q_1,\ldots,p_j,q_2,-q_2,\ldots)
\right.\nl && \left.
+ \sum_i 
 \hat{R}_{n+4}(\ldots,p_i,q_1,-q_2,q_2,-q_1,\ldots)\right]
\nl &
\times&
 \left\{\begin{array}{ll}
 \displaystyle  4/3,& \sgn~q_1^0 = \sgn~q_2^0, \\
 \displaystyle  2/3,& \sgn~q_1^0 = -\sgn~q_2^0. \\
 \end{array} \right. \label{R2loopcolor}
\ea
This formula turns out to be, formally, an exact identity.
In section \ref{sec:R2loopplanar} we will deduce it from \nr{R2loop} and
discuss some subtleties it involves.

\begin{figure}
\begin{center}
\includegraphics[width=10cm]{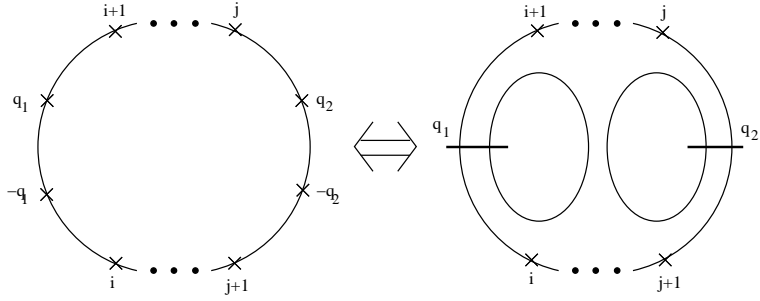}
\end{center}
\caption{Color structures for forward amplitudes are in
 one-to-one correspondence with cut planar diagrams in
 open-string theory.
}
\label{fig:openstring}
\end{figure}

A useful mnemonic for \nr{R2loopcolor} is that it is precisely
\nr{R2loop} applied
to planar open-string theory diagrams, as in Fig.~\ref{fig:openstring};
a similar version is obtained at 3-loop following section \ref{sec:3loop}.
This is quite pleasing: it simply means that the (space-time)
Schwinger-Keldysh formalism, which will be used below to derive \nr{R2loop},
is applicable directly at the level of individual open-string diagrams.

\subsection{Non-planar theories}

Equation \nr{R2loop}, viewed as the sum of a classical and a quantum part,
gives control over the quantum part.  We can try to determine to what
extent it is sensitive to unphysical data.
In section \nr{sec:uniqueness} we will argue that its integral
(the part with the $\frac13$'s) is in fact fully determined by
physical 3-particle cuts (only, at complexified momenta!)
This is not unreasonable, since the only reason this term had to be introduced was
to fix 3-particle cuts.  This term vanishes for non-overlapping loops, which have
no 3-particle cuts.
This leads us to make the following suggestion: the quantum part
may be implicitly determined by the classical part.
Lacking a good physical interpretation of such a result,
we will be content with formulating a conjecture.

This is best stated as a uniqueness conjecture.
Specifically, suppose one has a representation
\be
 R_n^{\ell-\textrm{loop}} = \sum_i C_n^i \times I_n^{\ell-\textrm{loop}},   \label{cuts1}
\ee
as a sum over ordinary $\ell$-loop Feynman integrals $I_n$ (e.g. any product of $1/(\sum q_i+\sum p)^2$-type
denominators together with numerator factors),
which are to be computed in Euclidean
space as usual and then continued to give the retarded amplitude.%
\footnote{
 Actually the same Euclidean amplitude, that is the same coefficients $C_n^i$,
 will determine both the retarded and time-ordered functions.
 The relevant analytic continuation is described in the next section.
}
The coefficients $C_n^i$ are rational functions of the external momenta only.
This is the same sort of expansion as is familiar from unitarity-based
methods, where one would take the $I_n$'s to span a
basis of scalar integrals such as
tadpoles, bubbles, triangles, boxes at 1-loop.
For the present discussion it will not matter what the $\ell$-loop basis is;
it could be complete or over-complete.
Since both sides are Feynman integrals one can take their classical
``$\ell$-particle cuts'', defined as the integrand to \nr{NAIVE}, and compare them:
\ba
 \lefteqn{\sum_{h_1,\ldots,h_\ell}(-)^{2h_1{+}\ldots+2h_\ell}
 R_{n+4}^\textrm{tree}(q_1^{h_1},-q_1^{-h_1},\ldots,q_\ell^{h_\ell},-q_\ell^{-h_\ell})}
&&\nl
&=& \sum_i C_n^i \times [\mbox{classical }\ell\mbox{-particle cuts]}\left(
   I_n^{\ell-\textrm{loop}}(q_1,-q_1,\ldots,q_\ell,-q_\ell)\right).
 \label{cuts2}
\ea
In general, \nr{cuts1} does not imply \nr{cuts2} since the right-hand side of
\nr{cuts1} is defined only modulo terms which integrate to zero.
Our uniqueness conjecture (supported in section \ref{sec:uniqueness} for $\ell\leq 3$) is the other implication:
\be
\noindent \mbox{
  {\bf Conjecture}. If \nr{cuts2} is an equality, then \nr{cuts1} is an equality.} \label{uniqueness}
\ee
The left-hand side of \nr{cuts2} is given by
the forward limit of physical tree diagrams, and can thus, in principle, be computed directly.
That means that if an Ansatz \nr{cuts1} for an integrand agrees with physical trees on all single-cuts,
then its integral gives the correct loop integral.
This gives a set of purely algebraic constraints to determine the basis coefficients $C_n^i$.

Some clarifications are in order.  First, any loop integrand is at best defined only modulo shifts
in loop momenta
(except in a planar theory, where origins can be defined,
but since we are now considering non-planar theories this cannot be invoked),
and the right-hand side of \nr{cuts1} is subject to this freedom.  The cut propagators, however, provide a
natural origin for the $\ell$ loop momenta and the comparison in \nr{cuts2} does not involve any choice
of origin (symmetrization under permutations of the loop momenta is implied).

Second, this allows the determination of loop integrands modulo integrands
with vanishing $\ell$-particle cuts.
At 1-loop, we believe that any such integrand is either a polynomial or a massless bubble.%
\footnote{
 The cuts of massless bubble (a loop integral with two internal propagator with no mass
 and a massless external momenta)  are equivalent to
 $1/(q{+}p)^2 + 1/(q{-}p)^2= 1/2q{\cdot}p + 1/-2q{\cdot}p=0$ because arbitrary shifts
 in the integration momenta are allowed.
 This is never a problem in practice, because massless bubbles integrate to zero
 in dimensional regularization.
}
More generally, the classification of such integrands is not known but we did not need it.
The conjecture is that any such integrand integrates to zero.

Third, there exists integrands which integrate to zero but which have non-vanishing single-cuts.
We will see an explicit example in section \ref{sec:5pt}; in general, a class of examples
are integrals which vanish upon symmetrical integration.
Because there is no integration in \nr{cuts2}, such terms do not necessarily vanish in it.
However it is also clear that one could perform e.g. symmetrical integration at the level
of \nr{cuts2}, if one wished.
Thus one is free to keep or ignore such terms as one wishes.
To put it differently, what is to be judged ``equal'' in \nr{cuts2}
depends on what one wants to achieve.
Similarly, whether the basis in \nr{cuts1} has to be a full algebraic basis,
as in the OPP method \cite{OPP}, or a more reduced basis, as in the conventional unitarity-based
methods, depends on the amount of vanishing integrals one wishes to keep.

Finally, it is important to note that matching all $\ell$-particle cuts on the two sides,
automatically implies that all unitarity cuts match.
It is certainly always possible to match all $\ell$-particle cuts.
Thus the conjecture would be proved, if a proof that ($D$-dimensional) unitarity cuts
suffice to determine any $\ell$-loop amplitude, existed for any theory.

\section{Response functions and Schwinger-Keldysh formalism}
\label{sec:keldysh}

We review
the basic properties of retarded response functions and their computation
within the Schwinger-Keldysh formalism.
Our presentation follows a semi-classical expansion which may be more familiar
from quantum mechanics.

The response functions are important for the present considerations, as they
dramatically simplify the physics and help keep equations such as \nr{R2loop}
intelligible.
The reader familiar with them could skip directly to the next section;
some useful references for this section are \cite{Chou,eijck,mueller}.

\subsection{Response functions}

Response functions describe the ``response'' to a perturbation.
One considers a system initially in its vacuum state $|0\rangle$
and add a  perturbation  $\mathcal{H}\to\mathcal{H}+ f(x)\OO(x)$ to its Hamiltonian density.
At time $t$ the system will be in the state
\be
  |\psi(t)\rangle_f = \mathcal{T}e^{-i\int^t d^Dx f(x)\OO(x)} |0\rangle,
\ee
where $\mathcal{T}$ is the time-ordering symbol and $\OO(x)$ is in the Heisenberg representation,
as usual in field theory.
By analogy with classical response functions,
quantum response functions should describe the expectation value of an
operator $\OO_1(x)$ as a function of $f$, or, more precisely, as a series expansion
in powers of $f$.
This motivates the definition for the $n$-point response function $R_n(1,2,\ldots n)$:
%\be
%   \langle \psi(t)| \OO_1(x_1) |\psi(t)\rangle_f \label{genfunc},
%\ee
\be
 R_n(x_1,x_2,\ldots x_n) \equiv
 \left.\frac{-(i)^{n}\delta^{n-1}}{\delta f(x_2) \cdots \delta f(x_n)}
   \langle \psi(t)| \OO_1(x_1) |\psi(t)\rangle_f
 \right|_{f=0}.  \label{genfunc}
\ee
It is possible to expand this definition in terms of vacuum expectation values of operator products.
The 2-point function becomes the usual retarded product
\be
 R_2(x_1,x_2)= -i \theta(x_1^0{-}x_2^0) ~\langle 0| [\OO_1(x_1),\OO_2(x_2)] |0\rangle,
\nonumber
% \label{retprop}
\ee
while $n$-point functions are expressed in terms of series of nested commutators
\cite{Chou}.  We shall not need such formulas.
%\begin{align}
% R_n(x_1; x_2,\ldots,x_n) = \sum_{\sigma\in S_{n{-}1}}
%\theta(x_1^0{-}x_{\sigma(2)}^0)
%\theta(x_{\sigma(2)}^0{-}x_{\sigma(3)}^0)
%\cdots \theta(x_{\sigma(n{-}1)}^0{-}x_{\sigma(n)}^0)
%\nl
%\times -i\langle0| [\cdots [[ \OO(x_1),\OO(x_{\sigma(2)})],
%\OO(x_{\sigma(3)})],
%\cdots],
%\OO(x_{\sigma(n)})] |0\rangle, \nonumber
%\end{align}
%in which the sum is over permutations of the $(n{-}1)$ indices $2\ldots n$.
%(For fermions, there is an additional minus signs for every permutation
%that exchange two fermions, in the usual way, so that e.g. the fermionic
%2-point function involves an anticommutator.)

\begin{figure}
\begin{center}
\includegraphics[width=14cm]{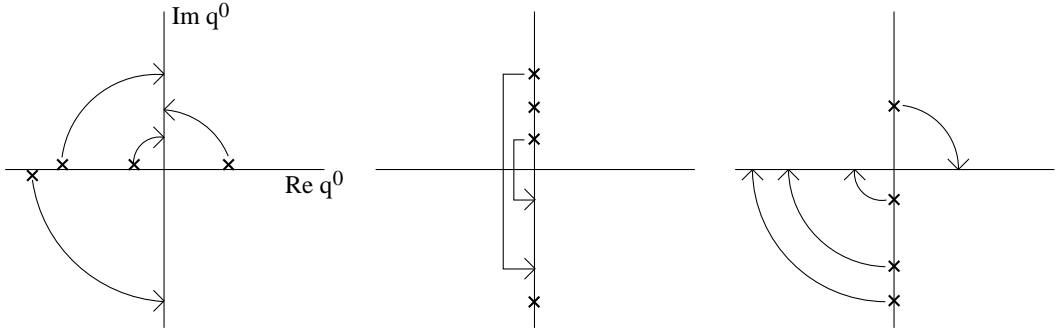}
\end{center}
\caption{The three-step analytic continuation which takes retarded
 amplitudes to time-ordered ones: from Minkowski (real $q^0$) to Euclidean
 signature, then in Euclidean signature from the positive to the
 negative imaginary axis, then from Euclidean to Minkowski signature
 through a Wick rotation.
}
\label{fig:cont}
\end{figure}

Important properties of these objects are:

\begin{itemize}
\item
They are Lorentz-covariant despite the explicit appearance of time
in the $\theta$-functions.
For the two-point function this may be seen from the vanishing
of commutators at space-like separated points.
More generally, this follows from the definition \nr{genfunc}:
imposing the vacuum state
at past infinity is a Lorentz-invariant boundary condition.

\item
In momentum space the response functions
are a simple analytic continuation
of the Euclidean amplitudes.  In this continuation,
a positive imaginary part is added to all incoming momenta
$p_2\ldots p_n$.
(The relative magnitudes of the imaginary parts is not important, because all discontinuities
 lie on the real axis.)
This is discussed in details in \cite{Evans}.
In particular, in vacuum, response functions are an analytic continuation
of time-ordered functions, since the latter are also obtained from Euclidean ones
by a Wick rotation.
 (The relation between response and Euclidean
 functions does extend to general, finite temperature, contexts \cite{Evans}.
 In that context retarded functions become the simplest objects,
 with time-ordered functions becoming neither particularly simple nor useful.)
For e.g. a simple power law the relationship is just
\be
   (p^2-i\epsilon)^{-\alpha}|_{\mbox{time-ordered}}
\Leftrightarrow
 (p^2 - i\epsilon p^0)^{-\alpha}|_{\mbox{retarded}}.\nonumber
\ee
In the general case, the analytic continuation represented in Fig.~\ref{fig:cont} should be used.
An explicit example will be given in the next subsection.
In general, it establishes that retarded and time-ordered amplitudes
have the same information content.%
%\footnote{For scattering amplitudes it is not possible
%to continue the individual momenta as above,
%because of the mass-shell constraints.  Presumably, a continuation of
%the relevant kinematic invariants should do the trick instead, but
%we have not investigated this in detail.
%For massless theories, a continuation
%starting as $q_i^\mu\to q_i^\mu (1+i\epsilon\sgn q_i^0)$
%for each incoming momentum $q_i$ will suffice.
%This keeps $q^2=0$ for all particles, while giving all
%invariants the correct time-like imaginary parts.  There is a constraint that
%the outgoing momentum remains null.
%}

\end{itemize}

\subsection{Example: a scalar triangle}

To help reassure
the reader that the integrals we discuss are
indeed closely related to familiar loop amplitudes,
let us immediately describe a simple but nontrivial example:
\be
 I_3=\int_q \frac{-i}{q^2(q{+}p_1)^2(q{-}p_2)^2}.  \label{triangle1}
\ee
Here we consider the case that $p_1^2=p_2^2=0$ for simplicity.
The usual result, with time-ordered boundary conditions,
is
\be
  I_3 = \frac{c_\Gamma}{\epsilon^2} (p_3^2-i\epsilon)^{-1{-}\epsilon} \label{triangleres}
\ee
where $c_\Gamma$ is a function of $\epsilon$ defined in Appendix \ref{app:triangle}
and $p_3= {-}p_1{-}p_2$.
Let us verify that \nr{1loop} indeed produces a simple analytic continuation of that answer.

\begin{figure}
\begin{center}
\includegraphics[width=13cm]{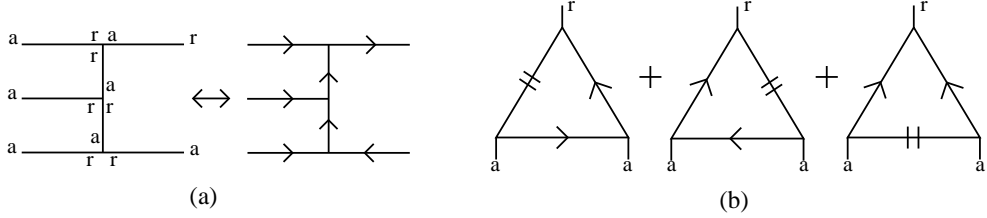}
\end{center}
\caption{Examples of tree and one-loop Schwinger-Keldysh diagrams
 for response functions.
}
\label{fig:ex}
\end{figure}

According to the latter formula, the triangle integral should be equivalent to
\be
 \int_q \pi\delta(q^2)\left[
 \frac{1}{(2q{\cdot}p_1 {+} p_1^2)
          (-2q{\cdot}p_2 {+} p_2^2)}
{+} \frac{1}{(2q{\cdot}p_2 {+} p_2^2)
          (-2q{\cdot}p_3 {+} p_3^2)}
{+} \frac{1}{(2q{\cdot}p_3 {+} p_3^2)
          (-2q{\cdot}p_1 {+} p_1^2)}\right].
% \nonumber 
\label{triangle2}
\ee
This sum is depicted in Fig.~\ref{fig:ex} (b).
The $i\epsilon$ prescriptions are also specified by the formula.
With retarded boundary conditions
one must choose one special leg, corresponding to the ``advanced'' one,
which we will take to be $p_3$.
Then in all terms, time is to flow from the cut propagator toward $p_3$.
For instance, the first term is to be read as
\be
 \frac{1}{(2q{\cdot}p_1 {+} p_1^2 -i\epsilon (p_1^0{+}q^0))
          (-2q{\cdot}p_2 {+} p_2^2 -i\epsilon (p_2^0{-}q^0))}. \nonumber
\ee
(A convenient mnemonic device, which works for all terms,
 is to generate the $i\epsilon$'s by adding small
 positive imaginary parts to $p_1^0$ and $p_2^0$.)
As a rule, we omit the explicit writing of $i\epsilon$ factors.

It is important to note that the $i\epsilon$'s are only needed when a retarded
propagator gets close to its mass shell.  The sign of its imaginary part,
opposite to the sign of its energy,
is then a well-defined Lorentz-invariant quantity.

The integrals are easily computed in a frame in which $p_1$ and $p_2$ are
within the $z-t$ plane, so that $2p_1{\cdot}p_2=-p_1^+p_2^-$
with $p_i^\pm\equiv p_i^0\pm p_i^z$.
We begin with the second term in the bracket.
Upon using the $\delta$-function to perform the $q_\perp$ integration in $D$ dimensions
it is reduced to
\be
% \int_q
%\frac{\pi\delta(q^2)}{(2q{\cdot}p_2 {+} p_2^2)
%          (-2q{\cdot}p_3 {+} p_3^2)}
% &=&
\frac{1}{(4\pi)^{D/2} \Gamma(1-\epsilon)}
 \int \frac{dq^+ dq^- \theta(q^+q^-)}{2(q^+q^-)^\epsilon}
 \frac{1}{q^+p_2^-(q^-p_1^+ + q^+p_2^- + p_1^+p_2^- + i\epsilon(q^0{+}p_1^0{+}p_2^0))}.
 \label{dummytri}
\ee
Three possible cases must be distinguished.

When both $p_1^+$ and $p_2^-$ are positive the pole occurs at negative
$q^+,q^-$.  The sign of the $i\epsilon$ prescription is dominated by that of $(p_1{+}p_2)^0>0$
(in the rest frame of $p_3$, $q^0$ carries half the energy),
so the pole is below the real axis.
Thus the integral for negative $q^+,q^-$ may be evaluated by rotating the two variables
by $180^o$ clockwise, while that for positive $q^+,q^-$ is done directly, yielding for \nr{dummytri} 
\be
%   \int \frac{dq^+ dq^- \theta(q^+q^-)}{2(q^+q^-)^\epsilon}
% \frac{1}{q^+(q^+ q^- + 1)}
 = \frac{\Gamma(1-\epsilon)\Gamma(2\epsilon)}{\epsilon (4\pi)^{D/2}p_3^2} 
  \times |p_3^2|^{-\epsilon} \frac{ 1 + e^{2\pi i \epsilon}}{2}.
\nonumber
\ee

When both $p_1^+$ and $p_2^-$ are negative one finds the complex conjugate
\be
\frac{\Gamma(1-\epsilon)\Gamma(2\epsilon)}{\epsilon (4\pi)^{D/2}p_3^2} 
  \times |p_3^2|^{-\epsilon} \frac{ 1 + e^{-2\pi i \epsilon}}{2}.
\nonumber
%-\frac{\Gamma(1-\epsilon)^2\Gamma(2\epsilon)}{\epsilon} \times \frac{ 1 + e^{2\pi i \epsilon}}{2}.
\ee

When $p_1^+$ and $p_2^-$ have opposite sign, that is $p_3^2$ is spacelike,
%the integral can be written
%\be
%\frac{1}{(4\pi)^{D/2} \Gamma(1-\epsilon)}
% \int \frac{dq^+ dq^- \theta(q^+q^-)}{2(q^+q^-)^\epsilon}
% \frac{-1}{q^+|p_2^-|(q^-|p_1^+| - q^+|p_2^-| + |p_1^+||p_2^-|)}.
% \nonumber
%\ee
the sign of the $i\epsilon$ addition depends on whether it is $q^+$ or $q^-$ which is
larger at the pole, which is determined by the sign of $q^+,q^-$.  For instance if $p_1^+>0$ then
for one sign the integral can be done by rotating $q^+$ clockwise and for the other sign
it can be done by rotating $q^-$ counter-clockwise.  The result is therefore
\be
\frac{\Gamma(1-\epsilon)\Gamma(2\epsilon)}{\epsilon (4\pi)^{D/2}p_3^2} 
  \times |p_3^2|^{-\epsilon} \frac{ e^{\pi i \epsilon} + e^{-\pi i \epsilon}}{2}.
\nonumber
\ee

The third term in the bracket \nr{triangle2}
gives exactly the same result and the first term gives zero, there being no scale in it.
To obtain the final answer the above results must thus simply be multiplied by 2.
The three different cases nicely combine into a single formula,
\be
 \frac{c_\Gamma}{\epsilon^2} \times (p_3^2 + i\epsilon p_3^0)^{-1{-}\epsilon}, \nonumber
\ee
which is precisely the claimed simple analytic continuation of \nr{triangleres}.

In practice, computing any one of these cases is enough, as all three are related through well-defined
analytic continuations.

If we had used a different retarded
boundary condition so that, for instance, $p_1$ or $p_2$ were the
advanced leg instead of $p_3$, the same computation would have landed us in a different branch of the result
\be
 \frac{c_\Gamma}{\epsilon^2} \times (p_3^2 - i\epsilon p_3^0)^{-1{-}\epsilon}, \nonumber
\ee
again in complete agreement with the analysis of the preceding subsection.
% It is trivial to see that the complex conjugate is obtained, there is no need to compute anything.

In summary, regardless of how results are obtained, it is trivial to convert
between time-ordered and retarded boundary conditions.
In the Euclidean region (spacelike invariants)
there is in fact no continuation at all to be done.

The reader interested in an example with a less trivial analytic form than \nr{triangleres}
is referred to Appendix \ref{app:triangle}.  There we compute the triangle
for general external momenta, which in $D$ dimensions is
expressed in terms of hypergeometric functions.
Full agreement is again obtained, which follows from a non-trivial identity between Feynman parameter
integrals of different dimensionality.

\subsection{Semi-classical limit using the Schwinger-Keldysh formalism}

Response functions admit a path-integral representation known
as the Schwinger-Keldysh path-integral.
This formalism is employed to study finite-temperature systems, and even
out-of-equilibrium systems.
Here, we use it to study a theory in its vacuum state.
But since the formalism is the same, it is worth noting
that much in this section (and paper) should generalize to these other cases.

This path integral runs over a doubled set of fields.
It would be impossible not to double the fields to compute \nr{genfunc}:
one set is needed to evolve the ket, and another is needed to evolve the bra.
Those fields are just integrations variables in a path integral.
Labeling them collectively as $\phi^1$ and $\phi^2$,
the full path integral is written as
\be
 \int \DD\phi^1\DD\phi^2 e^{iS[\phi^1]-iS[\phi^2]}. \nonumber
\ee

In the semi-classical regime the spread of the
wavefunction should be small,
heuristically $(\phi^1{-}\phi^2) \sim \hbar(\phi^1{+}\phi^2)$.
We expect this to be connected with the ordinary perturbative regime.
To exploit this we make a linear change of variables to the
so-called Keldysh basis consisting of averaged and differenced fields:
\be
 \phi^r\equiv \frac12 (\phi^1+\phi^2),\quad
 \phi^a\equiv \phi^1-\phi^2. \label{keldysh}
\ee
(We follow the notations used by the finite temperature field theory community.)
Now we expect simply $\phi^a\sim \hbar$.

The first step is to express the response functions in this basis.
As simple computation shows \cite{Chou},%
\footnote{When writing $\OO^a$ for a composite
 operator, the combination $\OO^1{-}\OO^2$ should be understood. 
 Thus, for instance, for $\OO=(\phi)^2$, we let $\OO^a\equiv (\phi^1)^2- (\phi^2)^2$
 which should not be confused with $(\phi^a)^2$.
 The same goes for $\OO^r\equiv \frac12(\OO^1{+}\OO^2)$.
}
\ba
 R_n(x_1,x_2,\ldots,x_n) &\equiv& -i \langle
  \OO^r(x_1) \OO^a(x_2)\cdots\OO^a(x_n) \rangle \label{correla}
\ea
if $\OO(x_1)$ is the advanced leg.
Notice that the perturbations $f\OO$ have become the differenced operators $\OO^a$,
which happens because it enters with opposite signs on the two branches of the contour.

The next step is to expand the action in powers of $\phi^a$.
This contains only odd powers,
the full action $S[\phi^1]-S[\phi^2]$ being
odd under interchange of the two fields.
In the leading, linear approximation, this becomes $\phi^a(\frac{\delta S}{\delta \phi}[\phi^r]-f)$
and the $\phi^a$ integration can be done exactly: it enforces
the Euler-Lagrange equations
\be
 \int \DD\phi^a e^{i\phi^a(\frac{\delta S}{\delta \phi}[\phi^r]-f)}
=
 \delta\left(\frac{\delta S}{\delta \phi(x)} [\phi^r] -f\right).
 \label{local}
\ee

Hence, at the leading order,
response functions are computed by solving
classical equations of motion with sources and retarded boundary conditions, as expected.
What are the corrections to \nr{local}?
They arise from
neglected terms in the full action, which are of order $(\phi^a)^3$.
Using \nr{correla}, their
effect on some $R_n$ is seen to correspond
to a response function $R_{n+3}$ with three legs
at the same space-time point, where an additional interaction vertex is inserted.
This has the topology of a 2-loop diagram.
Hence, the localization \nr{local} remains valid at 1-loop!
This is a remarkable and far-reaching statement,
yet it was to be fully expected from the estimate $\phi^a \sim \hbar$.

Equation \nr{local} being valid at 1-loop means that 1-loop corrections
only arise through initial conditions.
The path integral reduces to
a sum over classical solutions, each of which is determined by its own initial conditions,
but there may be a distribution of such initial conditions.
Thus our problem is one in classical statistical field theory,
and, indeed, \nr{local} looks very similar to the Siggia-Martin-Rose formulation \cite{Martin}
of statistical field theory.
At 1-loop it is sufficient to know only the 2-point function of the distribution, which is $\sim \hbar$.
The relevant 2-point function is that of the averaged field $\phi^r$ in vacuum,
which may be readily computed.  For a scalar field
\be
 G^{rr}(q) = \pi\delta(q^2+m^2). \label{stupidrr}
\ee
Inserting this 2-point function into an amputated response function
with $n{+}2$ legs will give the 1-loop response function.
For a theory with a real spinless scalar, with proper symmetry factors
this procedure immediately gives
\be
 R^\textrm{1-loop}_n(p_1,\ldots,p_n) = \frac12
 \int_q \pi\delta(q^2{+}m^2_h)~ R^\textrm{tree}_{n+2}(p_1,\ldots,p_n;q,-q).
 \label{1loopA}
\ee
Putting back spin indices, this precisely reproduces \nr{1loop},
providing a very simple and physically transparent proof of that result.

\subsection{Perturbation theory}

Feynman rules in the Keldysh basis are summarized in 
Appendix \ref{app:rules}.
They are not too complicated.
For concreteness, here we give the propagator which is a $2\times 2$ matrix,
\be
  \left(\begin{array}{ll}
  G^{rr} & G^{ra} \\ G^{ar} & G^{aa}\end{array}\right)
 = \left(\begin{array}{ll}
  \frac12 (G^>+G^<) & \GR \\ \GA & 0 \end{array} \right)
 \simeq \left(\begin{array}{ll}
  \pi\delta(p^2+m^2) & \frac{-1}{p^2-i\epsilon p^0} \\ \frac{-1}{p^2+i\epsilon p^0}
  & 0 \end{array} \right),
\nonumber
\ee

The key feature of this propagator is that $G_{aa}=0$ (which is a non-perturbative statement
known as a ``largest-time equation'': $\phi^a\to 0$ at large times.)
Furthermore, all vertices contain odd numbers of $a$ fields, and the most ``common''
vertices have only one $a$ field.  
In practice, this means that almost all propagators in a retarded amplitude will be retarded propagators.
Heuristically, vertices with more than one $a$ are suppressed by powers
of $\hbar^2$.

These heuristic estimates can be justified by rigorous counting.
Let $P$ denote the number of propagators in a diagram,
let $V$ be the number of vertices and let $n_a$ be the number of extra $a$ fields
at the internal vertices ($n_a$ is twice the number of $\hbar^2$-vertices).
Then, balancing the number of $a$ fields between propagators, vertices and
external legs (taking $n-1$ external legs to be
of $a$-type for a response function), gives
%\be
$ P-P_{rr} = n_a + V + n-1$,
%, \nonumber
%\ee
where $P_{rr}$ is the number of $rr$ propagators.
Combining this with Euler's formula for the loop number, $\ell= P -V +1-n$,
yields
\be
  \ell = P_{rr} + n_a.  \label{loopcount}
\ee
This is the desired formula.  Both terms are non-negative and $n_a$ is even.
Equivalently, $G^{rr}\sim \hbar$ and $\phi^a\sim \hbar$.\footnote{
 For a correlation function involving $n_r$ external $r$-fields,
 the left-hand side becomes $(\ell+n_r-1)$.}

Thanks to \nr{loopcount}
it is a simple exercise to use the Feynman rules of Appendix \ref{app:rules} to reproduce the
tree- and 1-loop results obtained above.  We leave this exercise to the reader.
At tree-level, one should find a sum over ordinary Feynman graphs but with all propagators
being retarded, since \nr{loopcount} says that there cannot be any cut propagator.
At 1-loop, one should get a sum over diagrams that have exactly
one cut propagator, as for instance in Fig.~\ref{fig:ex} (b), whose combinatorics are trivially shown to
reproduce \nr{1loop}.

\section{Formal manipulations of higher-loop integrals}
\label{sec:newres}

In this section we derive the results announced in section \nr{sec:guesswork}.
We begin by systematically analyzing 2-loop Schwinger-Keldysh diagrams,
which will lead to the formula \nr{R2loop} involving in general an unphysical
sum over tree diagrams, and to its 3-loop extension.

\subsection{Cutting 2-loop diagrams}

The new ingredient at 2-loop is the $\hbar^2$-vertex, that is the vertices with 3 $\phi_a$ fields.
Physically, this vertex is required for unitarity of the theory.
Without any loss in generality we can simplify
the combinatorics by
restricting to theories with interaction vertices
of at most cubic order:  Higher-order vertices could be
broken down to cubic ones at the cost of introducing auxiliary fields;
in the Schwinger-Keldysh formalism,
the propagator of an auxiliary field with kinetic term
$S=\frac12 \xi^2$ is simply
\be
  \left(\begin{array}{ll}
  G^{rr} & G^{ra} \\ G^{ar} & G_{aa}\end{array}\right)
 = \left(\begin{array}{ll}
  0 & i \\ i & 0 \end{array} \right).
\ee

\begin{figure}
\begin{center}
\includegraphics[width=10cm]{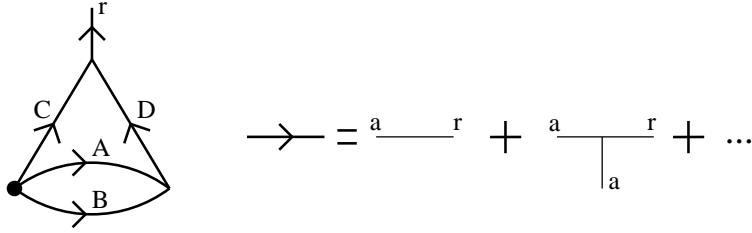}
\end{center}
\caption{Anatomy of a 2-loop diagram containing a $\hbar$-suppressed vertex,
shown as the dotted one.
The retarded propagators on the figure are really strings of retarded propagators
with insertions of external fields,
as illustrated on the right.
}
\label{fig:2loopX}
\end{figure}

The most general 2-loop
diagram with an $\hbar^2$-vertex has the form depicted in Fig.~\ref{fig:2loopX}.
The loops need to overlap, with is why this is no other possibility.
To simplify the discussion we omit the external $a$ fields,
which do not play any active role in this story -- their only effect is to
source a classical background field and ``dress'' the retarded propagators in the diagram.

In terms of dressed propagators, all diagrams reduce to Fig.~\ref{fig:2loopX}.
Let us dissect it in some detail.
We call the $\hbar^2$-vertex ``vertex 1''.
Starting from it and following the arrows there are three paths.
All three eventually merge together, but two of them will always
merge beforehand.  We label these two shorter segments A and B, ending
them at vertex 2 where they meet.
Then segment C is the third other one, going to the final point, and segment
D is the merger of segments A and B up to that same final point.
This labeling, shown in Fig.~\ref{fig:2loopX}, is unique up to the interchange of paths A and B.

The obvious symmetry
between vertices 1 and 2 suggests a special role for vertex 2.
Indeed, topologically speaking, in a 2-loop diagram with overlapping loops, there
are exactly two vertices which can be turned into $\hbar^2$-vertices.
These are those two for which all incoming momenta are free,
not restricted by the diagram's external momenta, and are vertex 1 and vertex 2.
In that sense they form a pair.

The propagators in the diagrams are all retarded. To cut-open two of them
we use causality as explained in the introduction.
Here the relevant identity stems from the definition of the retarded
propagator,
\be
 G_{R,A}(x{-}y) = \pm\rho(x{-}y)\theta(\pm(x^0{-}y^0)),
 \nonumber
% \label{GRdef}
\ee
where the spectral density
$
  \rho(x{-}y) \equiv [\phi(x),\phi(y)]
$
is supported, in momentum space, on the mass shell $q^2+m_h^2=0$.
Whenever a $\theta$-function of time is redundant we drop it and
gain a cut propagator.
This can be done at the level of ``dressed'' propagators, e.g.
strings of bare propagators,
using the telescopic identity
\be
 \GR\GR{\cdots}\GR %- \GA\GA\cdots\GA 
= \rho\GR{\cdots}\GR + \GA\rho{\cdots}\GR +\ldots
 +\GA\GA{\cdots}\rho, \label{mainid}
\ee
which holds modulo a term $\GA\GA\cdots\GA$
which vanishes when multiplied with a $\theta$-function of the time
separation between the string endpoints.
In words, each propagator in a cut string gets cut in turn.
Thus we can work with dressed propagators just like with ordinary propagators.

To apply this to the diagram, for instance, one can first cut-open propagators B,
then use the remaining $\theta$-functions along segments
C and (A+D) to cut-open a second segment.
This leaves a connected tree, and then there is no way to proceed
further using \nr{mainid}.

We shall need the most general identity obtained by this method and so let us
introduce a hyper-condensed
notation.  Let $G(A)$ denote the string of retarded propagators
of segment $A$ and let $\rho(A)$ is the right-hand side of
\nr{mainid} applied to it.
(We omit subscripts $G_{R,A}$ from the retarded propagators whenever
 the time flow is clear from the context -- time always flow from a cut propagator to its future.)
The most general identity consistent with symmetry between A and B then reads
\ba
 \GR(A)\GR(B)\GR(C)\GR(D) &=& \alpha~ G(A)\rho(B) \rho(C)G(D) 
\nl &+& 
\beta~\rho(B)G(C)\left[\rho(A)G(D) + \rho(D)G(A)\right]
\nl &+&   (A\leftrightarrow B), \nonumber
\ea
where $\alpha=\frac12-\beta$ is one free parameter.
We treat the diagrams where 2 is the $\hbar^2$ vertex in the same way,
with the same $\alpha$, so as to ensure symmetry under $1\leftrightarrow 2$.
Hence we have only one free parameter consistent with the symmetries.
On the other hand, two topologically distinct
cuts appear (A-B or A-C), and unique choice will make their coefficients equal.
This is our best hope at obtaining a ``natural'' formula.
This choice is $\alpha=\frac5{12}$ and $\beta=\frac1{12}$ leading to
\begin{align}
& \GR(A)\GR(B)\GR(C)\GR(D)
+\GA(A)\GA(B)\GR(C)\GR(D)
\nl
=& \frac13\left[
\rho(A)G(B)\rho(C)G(D) + G(A)\rho(B)\rho(C)G(D)
\right.\nl&\left.
-\rho(A)G(B)G(C)\rho(D) -G(A)\rho(B)G(C)\rho(D)
\right.\nl&\left.
+\rho(A)\rho(B)G(C)G(D) \right]. \nonumber
\end{align}

%\begin{figure}
%\begin{center}
%\includegraphics[width=12cm]{cutdiags.eps}
%\end{center}
%\caption{The terms in equation \nr{dres2}.
%}
%\label{fig:cuts}
%\end{figure}

The pattern of signs is the same as in \nr{R2loop}!
Using the same notation as there, and recalling factors of $\frac14$ from the vertices,
we have found that the sum of diagrams with a $\hbar^2$ vertex is
\ba
 R_n^\textrm{2-loop,$\hbar^2$ vertex}(p_1,\ldots,p_n) &=&
  \frac{1}{2!}
\prod_{i=1}^2\left[ \sum_{h_i} (-)^{2h_i}
\int_{q_i}
\pi\delta^+(q_i^2{+}m_{h_i}^2) \right]
\nl
&&\sum_\textrm{connected trees}
 R_{n+4}(p_1,\ldots,p_n;q_1^{h_1},-q_1^{-h_1},q_2^{h_2},-q_2^{-h_2})
\nl &
\times&
 \left\{\begin{array}{ll}
 \displaystyle  1/3,& \mbox{$q_1$ and $q_2$ parallel}, \\
 \displaystyle  -1/3,& \mbox{$q_1$ and $q_2$ anti-parallel}, \\
 \displaystyle  0,& \mbox{$q_1$ and $q_2$ non-comparable}.
 \end{array} \right. \label{R2loop_hbar}
\ea
Integrating out the auxiliary fields to restore original
four-point or higher vertices is easily shown to pose no problem in this form,
but this seems to be the case only for $\alpha=\frac5{12}$.
%(This would is only true for this special value of $\alpha$.)
Upon including the diagrams with no $\hbar^2$ vertices, which
are described by the classical contribution \nr{NAIVE},
this establishes the 2-loop formula \nr{R2loop}.

\subsection{3-loop formula}
\label{sec:3loop}

The 3-loop case is very easy to analyze once one realizes that
a 3-loop diagram contains at most one $\hbar^2$-vertex.
Those diagrams will contain one $rr$-propagator, which acts
like a classical fluctuation.  Stripping it off leaves
a 2-loop amplitude, which we have just analyzed.
More generally, the same analysis applies to any $\ell$-loop diagram with only
one $\hbar^2$ vertex.
One needs only be careful with symmetry factors.
For the sum of all 3-loop diagrams including those with no $\hbar^2$ vertex we obtain
\ba
 R_n^\textrm{3-loop}(p_1,\ldots,p_n) &=&
  \frac{1}{3!} %\sum_{h_1,h_2,h_3} (-)^{2h_1+2h_2+2h_3}
\prod_{i=1}^3\left[ \sum_{h_i} (-)^{2h_i}
\int_{q_i}
\pi\delta^+(q_i^2{+}m_{h_i}^2) \right]
%~\pi\delta^+(q_2^2{+}m_{h_2}^2)
%~\pi\delta^+(q_3^2{+}m_{h_3}^2)
\nl
&\times& \sum_\textrm{connected trees}
 R_{n+6}(p_1,\ldots,p_n;q_1^{h_1},-q_1^{-h_1},\ldots,q_3^{h_3},-q_3^{-h_3})
\nl &
\times&
 \left\{\begin{array}{ll}
 \displaystyle  2,& \mbox{$q_1$ and $q_2$ parallel}, \\
 \displaystyle  0,& \mbox{$q_1$ and $q_2$ anti-parallel}, \\
 \displaystyle  1,& \mbox{$q_1$ and $q_2$ non-comparable}.
 \end{array} \right. \label{R3loop}
\ea

At 4-loop level one will find diagrams that contain two
$\hbar^2$-vertices.  Their analysis will not be attempted here.

\subsection{Planar gauge theories}
\label{sec:R2loopplanar}

In section 2 we gave a formula for 2-loop corrections in planar gauge theories,
which we reproduce here
\ba
 \hat{R}_n^\textrm{2-loop,planar}(p_1,\ldots,p_n) &=&
  N_c^2
\prod_{i=1}^2 \left[ \sum_{h_i} (-)^{2h_i}
\int_{q_i}
\pi\delta^+(q_i^2{+}m_{h_i}^2) \right]
\nl &\times& \left[
 \sum_{i\leq j}
 \hat{R}_{n+4}(\ldots,p_i,q_1,-q_1,\ldots,p_j,q_2,-q_2,\ldots)
\right.\nl && \left.
+ \sum_i 
 \hat{R}_{n+4}(\ldots,p_i,q_1,-q_2,q_2,-q_1,\ldots)\right]
\nl &
\times&
 \left\{\begin{array}{ll}
 \displaystyle  4/3,& \sgn~q_1^0 = \sgn~q_2^0, \\
 \displaystyle  2/3,& \sgn~q_1^0 = -\sgn~q_2^0. \\
 \end{array} \right. \label{R2loopcolor1}
\ea
We now show that this Ansatz follows from \nr{R2loop},
which is true in any theory.

Indeed, \nr{R2loopcolor1} and \nr{R2loop} only differ in the treatment
of ``non-comparable'' Feynman diagrams, which are now made ``comparable''
by their color structures.
We have to show that this difference integrates to zero.

%(As noted in section 2, there may be issues in taking the forward limit
% of the amplitudes in \nr{R2loopcolor}.  Here we simply assume that the resolution
% of these issues will not interfere with the present argument.  Indeed it is hard to imagine
%  how a tadpole-like ambiguity could have a profound impact on the value of an amplitude.)

The relevant cancellation pattern is embodied in \nr{mainid}. It is a multi-particle generalization
of \nr{toy0}.
Consider first the disjoint color structure, the first case in the Ansatz.
In the ``non-comparable''
diagrams, by definition, the $q_1$ and $q_2$ loops are disjoint from each other.
(They may share a vertex,
 which will not interfere with the present argument, but they share no propagator.)
Each such Feynman diagrams will generate one term for each choice, in each loop, of which
propagator is cut.
The color flow and the momentum flow follow each other along the loop so
the sum is of the telescopic form \nr{mainid}.  But now we have a telescopic sum around a full circle,
so both $q_1$ and $q_2$ integrands reduce to products of retarded propagators.  They integrate to zero.

The ``nested'' color structure is treated similarly.  The Feynman diagrams in this class
that have two disjoint loops now simply have tadpole insertions; this class of diagrams
does not overlap with the ones just considered.

Equation \nr{R2loopcolor1} is still a rather formal result.
Difficulties appear in the forward limit $q'\to q$:
Tadpoles divergences
\be
 A(q,-q') - A(-q',q) \sim \frac{1}{(q-q')^2}, \quad q'\to q, \nonumber
\ee
no longer cancel, as the familiar cancellation in gauge theory requires a plus sign between the two terms.
In field theory parlance, the familiar cancellation of
$\sim f^{abc}$ tadpoles when contracted with $\delta^{ab}$ is spoiled by the combination \nr{R2loopcolor}.
This means that in planar gauge theories one trades the unphysical amplitude \nr{R2loop}
to the forward limit of a physical one, at the cost of a dependence on how the forward limit is taken.
The proper analysis of this ambiguity lies beyond the scope of this paper.
We mention, however, that for sufficiently supersymmetric planar theories the
tadpoles cancel in the susy sum and \nr{R2loopcolor1} is unambiguous as it stands!
This will be discussed further in section \ref{sec:susy}.
At the level of an individual diagram, the equation is tested in Appendix \ref{app:2loop}.

\subsection{Evidence for the uniqueness conjecture}
\label{sec:uniqueness}

Finally we provide evidence, actually sketch of a proof,
starting from \nr{R2loop_hbar} of the 2-loop uniqueness conjecture \nr{uniqueness}.
The setup now is a general non-planar quantum field theory.
We will argue that the integral of the quantum term \nr{R2loop_hbar} is fully determined
by 3-particle unitarity cuts, only, evaluated at complex loop momenta.
We will then argue that these cuts are physical and determined by their classical counterparts,
which together would imply that a loop integral with vanishing classical 2-particle cuts is zero.
%By energy positivity, the 3-particle cuts in the unphysical region
%where the energies have opposite signs is determined are exactly opposite to that classical contribution \nr{NAIVE},
%and by analyticity it will follow that they are the same everywhere.

The strategy is to employ contour integration in the rapidity plane
of one of the on-shell particles.
This requires space-time dimension $D\geq 2$.
When $D>2$, the choice of a rapidity plane breaks Lorentz invariance, and
the usefulness of the resulting formula for practical computations is unclear.
It suits, nonetheless, our present purpose.

The cancellation which will play a role is exemplified by the following integral:
\be
  \int_q\pi\delta(q^2)\sgn(q^0)\left[\GR(q{+}p) + \GA(q{-}p)\right] = 0. \label{toy0}
\ee
This integral vanishes by construction because, modulo shifts of $q$, it equals
the contour integral of $1/q^2(q{+}p)^2$ at infinity in the complex frequency plane.
The only reason the integrand in \nr{toy0} appears to be non-zero is the
relative shift between the two terms.
Really, the integrand is zero, this is just hard to see.
More precisely, the bracket times the sign function is zero when
viewed as a function on phase space and a certain relative shift between the two terms is made.
In any case, this vanishing will become more manifest once we perform
a second contour integration, which will break the shift symmetry.

\begin{figure}
\begin{center}
\includegraphics[width=6cm]{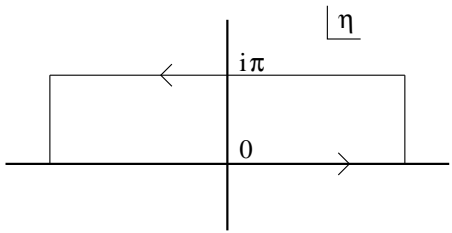}
\end{center}
\caption{Integration contour $C$ in the rapidity plane used in sectin \ref{sec:uniqueness}.  The interior
 of this contour is characterized by $\Im p_z>0$.
}
\label{fig:contourC}
\end{figure}

To do that second integration, the crucial point is to note that \nr{toy0}
is itself a contour integral in the phase space parametrized
by $q$.  This crucially relies on the sign function.
To see this, introduce a rapidity variable such that
$q^\mu = (\sqrt{q_\perp^2{+}m^2} \cosh \eta,$ $ \sqrt{q_\perp^2{+}m^2} \sinh \eta,q_\perp)$.
The integration measure becomes
\be
 \int_q \pi\delta(q^2)\sgn(q^0) = \int d^{D{-}2} q_\perp \oint_\mathcal{C} d\eta. \nonumber
\ee
The contour $\mathcal{C}$ has two real segments located at $\Im \eta=0,\pi$, as depicted in Fig~\ref{fig:contourC},
corresponding to positive and negative energies.
Crucially, the sign function provides the correct orientation for the contours.
The contour can be closed at infinity on each side
by adding small vertical segments, the integral over which vanishes.
Hence this integral is determined by the residues in its interior, which characterized by $\Im p_z>0$.
(This ``interior'' may not appear parity-invariant.
 However, the lines $\Im p_z=0$, on which the original contour runs, divide the complex phase-space into two halves
 which are parity conjugate to each other.
 Closing the contour on either half will give the same answer, which is thus parity-invariant.)

If the first term in \nr{toy0} has a pole at $q=\tilde q$, then the second term has a pole
at $\tilde q-p$ corresponding to the same physical configuration, e.g., the same unitarity
cut of a bubble integral.
If we assume that $\Im p_z=0$ (which we certainly can, for generic, off-shell external momenta),
the two poles are always inside or outside the contour at the same time.
The residue of the first pole is
\be
  \frac{1}{p^0\tilde q^z - p^z\tilde q^0}
%{\cdot}
% \left(  \begin{array}{l} \tilde q_z \\ \tilde q^0
%         \end{array}
%  \right)}
 .  \nonumber
\ee
The residue of the pole in the second term is
\be
  \frac{-1}{p^0(\tilde q^z{-}p^z) {-} p^z(\tilde q^0{-}p^0)}. \nonumber
\ee
Since the shift does not change the answer, the residues are equal and opposite
and we see that \nr{toy0} vanishes, as expected.

This cancellation pattern, for phase space integrals involving $\sgn(q^0)$, is all we need.
We apply the same technique to the $q_1$ integral in \nr{R2loop_hbar}.
Let $q_2$ be a cut propagator along, say, segment B (in the notation of Fig.~\nr{fig:2loopX}).
The two extra propagators cut by the $q_1$ contour integral can then be on AA, CC, DD, CD, AC or AD.
However, all but the last two cases AC and AD are easily seen to cancel by the above pattern.
The remaining cases AC and AD correspond, including cut propagator along B,
to genuine 3-particle cuts, e.g. the exchange of a 3-particles state, and the reason they do
not cancel is due to a relative sign introduced by the weighting in \nr{R2loop_hbar}.
The same conclusion is obtained regardless of where $q_2$ is in the diagram: only genuine 3-particle
cuts survive.

There remains the claim that the required 3-particle cuts are just ordinary unitarity cuts
and so are determined by (residues of) the forward limit of physical tree amplitudes.
It is hard to see how this could not be the case, given that
in the real but unphysical region considered around \nr{energypos},
perfect cancellation must occur between the amplitude needed here and a (residue of a)
physical forward amplitude.
Then by analytic continuation this should continue to hold at complex momenta.
It would be good to see this more explicitly.

\section{Example: 2-loop vacuum polarization}
\label{sec:ex}

To help make these considerations more concrete, let us now compute
the 2-loop vacuum polarization in massless QED
\be
 \PiR^{\mu\nu}(p) \equiv -i\int d^4x d^{-ip{\cdot}x}
    \langle0| [J^\mu(x),J^\nu(0)] |0\rangle \theta(x^0).
\ee
The current will be off-shell, $p^2\neq 0$.
This quantity was computed long ago by K\"all\'en and Sabry \cite{Kallen1955};
the massless limit enters the
the 3-loop electron contribution to the muon $g{-}2$.
It also appears in QCD,
where its imaginary part is related to the leading $\OO(\alpha_s)$
corrections to the total $e^+e^-\to \mbox{hadron}$ cross-section.
We will see here that the well-known result can be derived
using only on-shell data, even though we make no claim as to the efficiency
of this approach for this particular computation.

\subsection{1-loop vacuum polarization}

\begin{figure}
\begin{center}
\includegraphics[width=12cm]{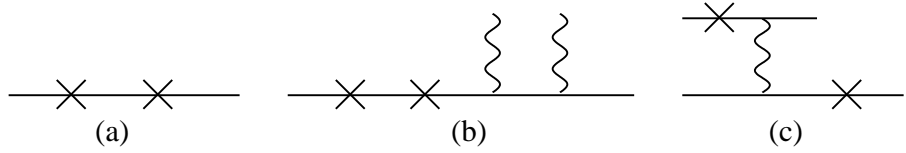}
\end{center}
\caption{Forward amplitudes against two off-shell currents, denoted as crosses,
 contributing to 1- and 2-loop vacuum polarization.  All external legs
 are on-shell.
}
\label{fig:2loop}
\end{figure}

Let us begin at 1-loop.  We need the forward amplitude of a quark
against two currents, as in Fig.~\ref{fig:2loop} (a).
Labeling the momentum of the incoming quark as $q$ the
forward amplitude is
\be
  \AA_1^{\mu\nu}(q) = e^2\Tr \left[ \qs \gamma^\mu \frac{\qs + \ps}{(q+p)^2} \gamma^\nu 
          + \qs \gamma^\nu \frac{\qs - \ps}{(q-p)^2} \gamma^\mu \right],
%\nl
%&=&
%  -4e^2 \frac{q^\mu(q{+}p)^\nu + q^\nu(q{+}p)^\mu - \eta^{\mu\nu}q{\cdot}p}{2q{\cdot p}+p^2} + (p\leftrightarrow -p).
 \label{forward}
\ee
where $q^2=0$.  
The helicity sum in \nr{1loop} simply adds the same thing with $q\to -q$.
We have absorbed
the minus sign $(-)^{2h}$ into \nr{forward},
so \nr{1loop} becomes
\be
  \PiR^{\mu\nu}(p) = \int_q \pi\delta(q^2) \AA_1^{\mu\nu}(q).
\ee

The integrand is manifestly transverse: $p_\mu\AA_1^{\mu\nu}(q)=0$.
Transverseness traditionally holds only for integrals,
but, in an on-shell formulation, it is already manifest
at the integrand level.
Lorentz-covariance is also manifest.
Transverseness and Lorentz invariance imply, in the standard fashion,
that we only need to compute the trace.
\ba
\PiR^{\mu\nu}(p) &=& (p^2\eta^{\mu\nu} - p^\mu p^\nu)
\int_q\pi\delta(q^2) \frac{\AA^\sigma_{1\,\sigma}(q)}{(D{-}1)p^2}
\nl
 &=& 2\frac{D{-}2}{D{-}1} e^2(p^2\eta^{\mu\nu} - p^\mu p^\nu) \int_q \pi\delta(q^2)
 \left[ \frac{1}{(q_1+p)^2} + \frac{1}{(q_1-p)^2}\right]\!. %-\frac{2}{p^2} \right].
 \label{int1}
\ea
In the bracket we have dropped a tadpole not depending on $p$,
which vanishes in dimensional regularization.

This $D$-dimensional integral may be readily carried out like the previous triangle example.
An equivalent, but much simpler, way to proceed is to recognize
\nr{int1} as the on-shell representation of a scalar bubble,
and evaluate the latter using standard Euclidean space techniques.
That is,
\ba
 \int_q \pi\delta(q^2) \left[ \frac{1}{(q+p)^2} + \frac{1}{(q-p)^2}\right] &=&
 \int_q \frac{-i}{q^2(q+p)^2}
\nl
 &=& \frac{\Gamma(\epsilon)}{(4\pi)^{\frac{D}{2}}} \frac{\Gamma(1-\epsilon)^2}{\Gamma(2-2\epsilon)}
 (p^2)^{-\epsilon}.
 \label{int1d}
\ea
The first equality holds because taking single-cuts of the right-hand side
yields the left-hand side.
(In Minkowski signature for the final result one has to remember the branch of the function $(p^2)^{-\epsilon}$
 appropriate for retarded boundary conditions, $p^2\to p^2-i\epsilon p^0$.)
This way the well-known $\MSbar$ result is reproduced:
\be
 \PiR^{\mu\nu}(p) = \frac{\alpha}{3\pi}
(p^2\eta^{\mu\nu} - p^\mu p^\nu)\left[-\log\frac{p^2}{\overline{\mu}^2} + \frac53\right],
\ee
where $\overline{\mu}^2=\mu^2e^{\log 4\pi-\ge}$ and $\alpha=e^2/4\pi$.

\subsection{2-loop vacuum polarization}

In the present example a new subtlety appears at 2-loop:
pinching singularities $\propto \delta(q^2)/q^2$
due to self-energy subdiagrams.  Indeed the formula
\nr{R2loop} contains no instruction to ``amputate,''
and, besides, it is hard to see how one could ``amputate'' diagrams in the present framework
without spoiling gauge invariance.

We postpone a more satisfactory treatment of these singularities to subsection \ref{sec:pinch}.
For the time being we just use a quick fix.
Thinking of the function $\delta(q^2)$ as the result of a contour integration
over a small circle around $1/q^2$, suggests viewing
$\delta(q^2)/q^2$ as a poor guess for a contour integral around $1/2q^4$.
Computing a residue around $1/2q^4$ ought to be the right thing to do.
This leads to the prescription, for any $F(q)$ regular at $q^2=0$,
\be
 \int \frac{\pi\delta(q^2)}{q^2} F(q) \to \frac{1}{2 i}\oint \frac{F(q)}{2(q^2-i\epsilon)^2}
  \simeq \frac12 \int -\pi\delta'(q^2) F(q) \label{pinch}.
\ee
The symmetry factor $1/2$ is because when two
propagators pinch in a diagram,  $\delta(q^2)/q^2$ gets generated once for each propagator.
Treating each propagator using \nr{pinch} will produce the right answer.
Using this prescription we can evaluate the forward amplitude
of a fermion against two currents plus two photon insertions,
depicted in Fig.~\ref{fig:2loop} (b) (summed over permutations), which contribute to
$\PiR^{\mu\nu}$ an amount
\ba
 && -\int \pi\delta'(q_1^2)\pi\delta(q_2^2) \Tr\left[
 q_1 \epsilon \frac{1}{q_1+q_2} \epsilon^* q_1
 \gamma^\mu \frac{1}{q_1+p}\gamma^\nu + ((p,\mu)\leftrightarrow (-p,\nu)) \right] \nl
&& \hspace{-1cm}
+\int \pi\delta(q_1^2)\pi\delta(q_2^2)\Tr\left[
 q_1 \epsilon \frac{1}{q_1+q_2} \gamma^\mu
 \frac{1}{q_1+q_2+p} \gamma^\nu \frac{1}{q_1+q_2} \epsilon^* \right.
\nl
&&\left. \hspace{3cm}
 + q_1\gamma^\mu \frac{1}{q_1+p} \epsilon
  \frac{1}{q_1+q_2+p} \gamma^\nu \frac{1}{q_1+q_2} \epsilon^* \right.
\nl
&&\left. \hspace{3cm}
 + q_1\gamma^\mu \frac{1}{q_1+p} \epsilon
  \frac{1}{q_1+q_2+p} \epsilon^* \frac{1}{q_1+p} \gamma^\nu \right.
\nl
&&\left. \hspace{3cm}
 + q_1 \epsilon \frac{1}{q_1+q_2} \gamma^\mu \frac{1}{q_1+q_2+p}
\epsilon^* \frac{1}{q_1+p} \gamma^\nu \right.
\nl
&&\left. \hspace{7cm}
+ ((p,\mu)\leftrightarrow (-p,\nu))\right]. \label{longstupid}
\ea
This is to be summed over photon polarizations $\epsilon$;
% transverseness p_\mu just checked.  Also for \epsilon\to q_2, except you need full (q_2,\epsilon)<->(-q_2,\epsilon^*)
% symmetry.
we will systematically use symmetry under $q\to -q$ to simplify
 the writing of the integrands.
Keeping this in mind and using the identity $q_1^2\delta'(q_1^2)=-\delta(q_1^2)$,
it is easy to check that \nr{longstupid}
is manifestly transverse under both $\epsilon^\alpha \to q_2^\alpha$
and $\gamma^\mu\to \pslash$.

It is important to stress that \nr{longstupid}
includes only physical polarizations $\epsilon,\epsilon^*$ for the exchanged photon.
No gauge fixing is needed because that photon is on-shell.
Longitudinal photon polarizations manifestly decouple at the level of the integrand.
This implies that the polarization sum is equivalent to a Lorentz trace
$\sum_\lambda \epsilon_\lambda^*{}^\alpha \epsilon_\lambda^\beta\to\eta^{\alpha\beta}$.
This fact will be used here, for convenience, as it helps simplify the evaluation of
the individual terms as opposed to their sums.

The trace of \nr{longstupid} is still a complicated function.
To avoid unimportant details, we now show details only for the terms having
one particular Dirac structure,
$\Tr[\aslash{A} \gamma^\mu \aslash{B} \gamma_\mu \aslash{C}\gamma^\alpha \aslash{D}\gamma_\alpha]$,
with obvious notation, corresponding to the first, second and fourth line of \nr{longstupid}.
The mass-shell delta-functions allow the numerators to be simplified
and $\PiR{}^\mu_\mu/e^4$ to be reduced to
\begin{align}
16(1-\epsilon)^2 \int_{q_1,q_2} \pi\delta'(q_1^2)\pi\delta(q_2^2)
 & \left[\frac{p^2}{(q_1{+}p)^2} -1 \right]
\nl
%+16(1-\epsilon)^2 \int_{q_1,q_2}\pi\delta(q_1^2)\pi\delta(q_2^2)
% &\left[ \frac{p^2}{(q_1{+}p)^4} + \frac{1}{(q_1{+}p)^2}
% \right. 
%\nl
% &\left.
 % + \frac{2p{\cdot}q_2}{(q_1{+}p)^2(q_1{+}q_2)^2}\right].
 %\nl
+16(1-\epsilon)^2 \int_{q_1,q_2}\pi\delta(q_1^2)\pi\delta(q_2^2)
 &\left[    \frac{-2p{\cdot}q_2}{(q_1{+}q_2)^2(q_1{+}p)^2}
   +\frac{p{\cdot}(p{+}2q_1{+}2q_2)}{(q_1{+}p)^2(q_1{+}q_2{+}p)^2}\right.
 \nl
   &\hspace{-2cm}
\left. +\frac{2p{\cdot}q_2}{(q_1{+}q_2)^2(q_1{+}q_2{+}p)^2}
  -\frac{p^2}{(q_1{+}p)^4} + \frac{1}{(q_1{+}q_2{+}p)^2}\right].
 \label{long1}
\end{align}

Next we need to compute the forward amplitudes of two fermions against
the currents as depicted in Fig.~\ref{fig:2loop} (c), which are
listed in the Appendix \ref{app:ferm}.
These are manifestly transverse by themselves
and again it suffices to compute their trace.
The terms that have the same Dirac structure as above are
\begin{align}
 16(1-\epsilon)^2 \int_{q_1,q_2} \pi\delta'(q_1^2)\pi\delta(q_2^2)
 &\left[1- \frac{p^2}{(q_1{+}p)^2}\right]
 \nl
 16(1-\epsilon)^2 \int_{q_1,q_2}\pi\delta(q_1^2)\pi\delta(q_2^2)
 &\left[
  \frac{2p{\cdot}(q_1{-}q_2)}{(q_1{-}q_2)^2(q_1{+}p)^2}
  + \frac{p{\cdot}(2q_2-p)}{(q_1{+}p)^2(q_1{+}p{-}q_2)^2} \right.
  \nl &\left.
   + \frac{p^2}{(q_1{+}p)^4}
  \right] \label{long2}
\end{align}
The cancellation of the $\delta'$ terms
between \nr{long1} and \nr{long2} is due to chiral symmetry, which forbids
a massless electron from acquiring a mass.

At this stage it would seem that we have many non-standard integrals to compute.
Furthermore, we don't even have the whole integrand yet since the non-standard
amplitudes with the $\frac13$'s in \nr{R2loop} are not yet included.

The beauty of the formalism is that this is unnecessary.
The on-shell integrals are guaranteed to sum up to the classical cuts
of standard Feynman integrals.
In the present case, it is easy to see that modulo shifts in the integration variables,
the sum of \nr{long1} and \nr{long2} is reproduced by the classical cuts of
\be
% \PiR{}^\mu_\mu(p) \supset
 16(1-\epsilon)^2 \int_{q_1,q_2} \frac{(-i)^2}{q_1^2q_2^2} \left[ 
 \frac{p{\cdot}q_1}{(q_1{+}q_2)^2 (q_1{+}p)^2} + \frac{1}{(q_1{+}q_2{+}p)^2}\right].
 \label{intscalar}
\ee
This is now a set of standard Feynman integrals.

It seems intuitively clear that the passage from \nr{long1} and \nr{long2} to \nr{intscalar} is unique,
modulo equivalent ways of rewriting the latter.
In general, this is the content of the conjecture \nr{uniqueness}.
In the present case there is actually nothing conjectural about it, since the exact same computation
could have been done in a planar gauge theory.

Repeating the same steps for the remaining Dirac structures in \nr{longstupid} and \nr{fermcut},
the forward amplitudes are found to uniquely determine the vacuum polarization as
\ba
 \Pi_R^{\mu\nu}(p) &=&
 8e^4 \frac{D{-}2}{D{-}1} \frac{p^2\eta^{\mu\nu}-p^\mu p^\nu}{p^2} \int_{q_1,q_2} \frac{(-i)^2}{q_1^2q_2^2}
 \left[ 
  \frac{-\frac12p^4}{(q_1{+}p)^2(q_2{+}p)^2(q_1{-}q_2)^2}
\right.
\nl &&\left.
+\frac{(\frac32 +\frac\epsilon2)p^2}{(q_1{+}q_2)^2(q_1{+}p)^2}
-\frac{(\frac34 + \frac\epsilon2) p^2}{(q_1{+}p)^2(q_2{+}p)^2}
 + \frac{ \frac12  \frac\epsilon2}{(q_1{+}q_2{+}p)^2}
\right]
\nl
%&=&
% 8\frac{D-2}{D-1} e^4 (p^2\eta^{\mu\nu}-p^\mu p^\nu) (p^2)^{-2\epsilon} c^2_\Gamma
% \left[ \frac{3}{8\epsilon} - 3\zeta(3) + \frac{89}{16}\right]
%\nl
&=&
\frac{\alpha}{3\pi} \times   \frac{3\alpha'}{4\pi}
 %\frac{3\alpha_s C_F}{4\pi}
\left[
 -\log\frac{p^2}{\overline{\mu}^2} -4\zeta(3) + \frac{29}{3}\right].
\label{result}
\ea
The first term may be readily integrated using
integration-by-parts techniques of \cite{IBP}
and the other integrals are straightforward.

Actually, the first line is just what one would have obtained
from computing ordinary Feynman diagrams, as it must.
The point of the present exercise was to deduce it directly from its cuts.

As an extra consistency check on the formalism, the scalar integral
corresponding to the second term in the bracket is evaluated
in its full on-shell form in Appendix \nr{app:2loop}, giving a nontrivial check
of the factor $\frac13$ in \nr{R2loop}.  The result agrees.

The imaginary part of \nr{result} gives, with the proper color factors,
the famous correction $[1+\frac{3\alpha_s}{4\pi}C_F]$ to the total
$e^+e^-$ hadronic cross-section in QCD.
We deduced it here starting from physically observable forward scattering amplitudes.

\subsection{Pinching poles and $\delta'(q^2)$ from forward limits}
\label{sec:pinch}

To complete our claim that \nr{result} comes from physical data,
it is important to check that the $\delta'(q^2)$ terms in \nr{longstupid}
does come from physical data.
This is not entirely obvious, because $q^2\delta'(q^2)\neq 0$ would suggest that the formula
actually needs information about infinitesimally off-shell particles.
We show here that the required data is contained within the forward limit of
a physical on-shell amplitude.

We consider the following on-shell, but non-forward, deformation of the tree amplitude:
\be
 R_{n{+}4}(\epsilon) =
 R_{n{+}4}(p-\epsilon q_2,-p;q_1,-q_1,q_2+\epsilon q_2,-q_2).
\nonumber
\ee
This conserves total momentum.  %Because it changes $q_2$ only by an overall rescaling,
%this formula does not introduce new arbitrary phases in the polarization tensors
%of $q_2$, which is convenient.
The limit $\epsilon\to 0$ has to be taken carefully
for the Feynman diagrams of self-energy type,
for instance Fig.\ref{fig:electron} (d) with $p\leftrightarrow q_2$.
In such diagrams, the two terms in which the cut fermion is on either side of the self-energy insertion
sum up to a $0/0$ form
\begin{align}
 A_\epsilon\approx \frac{1}{2\epsilon q_1{\cdot}q_2}
  \Tr&\left[ \qslash_1 \AA_L(q_1,q_2{+}\epsilon q_2,-q_2,-q_1{-}\epsilon q_2)  (\qslash_1{+}\epsilon \qslash_2) 
\right.\nl
 &\hspace{2cm}\left.
        \times  \AA_R(q_1+\epsilon q_2, p-\epsilon q_2,-p,-q_1)
\right.\nl
 &\left. - (\qslash_1{-}\epsilon \qslash_2) \AA_L(q_1{-}\epsilon q_2,q_2{+}\epsilon q_2,-q_2,-q_1) \qslash_1
\right.\nl
& \hspace{2cm}\left.
  \times \AA_R(q_1,p{-}\epsilon q_2,-p,-q_1{+}\epsilon q_2)\right]
  \nl
 \approx \frac{q_2^\mu}{2q_1{\cdot}q_2} \frac{\partial}{\partial q_1^\mu}
   \Tr&\left[ \qslash_1 \AA_L(q_1,q_2,-q_2,-q_1)  \qslash_1 
          \AA_R(q_1, p,-p,-q_1) \right]. \nonumber
\end{align}
Here $\mathcal{A}_L$ and $\mathcal{A}_R$ stand for the two ``self-energy'' factors
on the two sides of the singular propagator.
Now using the identity
\be
\int_{q_1,q_2}  \pi\delta(q_1^2)\pi\delta(q_2^2) \frac{q_2^\mu}{2q_1{\cdot}q_2}
   \frac{\partial}{\partial q_1^\mu} F(q_1,q_2)=
 -\int_{q_1,q_2} \pi\delta'(q_1^2)\pi\delta(q_2^2) F(q_1,q_2), \nonumber
\ee
we see that the limit $\epsilon\to 0$ exactly reproduces the first line in \nr{longstupid}.
This establishes that only physical data enters \nr{longstupid}.\footnote{
 Sometimes it may be useful to have a deformation preserving a mass-shell
 condition for all legs, for instance $p^2=0$.
 In this case one could consider a more general deformation such as
\be
 R_{\epsilon_1,\epsilon_2} = R(p+\epsilon_1 q_1 + \epsilon_2q_2,-p,q_1+\epsilon_1 q_1,-q_1,
   q_2+\epsilon_2 q_2,-q_2,\ldots)
\ee
where $\epsilon_1 = -(p{\cdot}q_2)/(p{\cdot}q_1)\epsilon_2$ and the dots
stand for an arbitrary number of other external states.
We find that this can also be related to \nr{pinch}, upon subtraction of certain gauge-invariant
terms involving the scale dependence $q_1^\mu \frac{\partial}{\partial q_1^\mu}$
of the (physical) on-shell self-energies.
}

%\section{The 1-loop $\beta$-function of Yang-Mills theory}

\section{Supersymmetry}
\label{sec:susy}

Supersymmetric theories are special.  On a somewhat more modest scale, this is also seen
when the present formalism is applied to them. 

%The first is the vanishing on-shell of self-energy subdiagrams (mass corrections)
%or massless particles, at the amplitude level.
%his simplifies the forward limits described in the previous subsection.
%The second simplification is the possibiity to use BCFW-type recursion
%relations to compute the tree amplitudes, due to the good ultraviolet behavior of the theory.
%We illustrate the formalism on a simple example in $\mathcal{N}=4$ SYM theory 
%in subsection \ref{sec:n4sym}.

\subsection{LSZ reduction}

The above-described formalism has been set up for the computation of off-shell correlation functions.
Ultimately, we want to consider on-shell S-matrix elements, which is where its full power
should become manifest.
To do so, the LSZ reduction procedure has to be adapted
to make it applicable to loop integrands.
As a first step toward this goal we now present how this works for sufficiently supersymmetric theories,
where supersymmetry cancellations simplify this task.
Here ``sufficient'' will turn out to mean $\mathcal{N}=1$ for massless particles
and $\mathcal{N}=2$ for massive particles.
Discussion of non-supersymmetric theories is left to future work.

To see the subtleties, consider an external gaugino leg, 
in $\mathcal{N}=1$ supersymmetry, whose momentum $p$ approaches its mass shell $p^2=0$.
In this limit the forward tree amplitude contains terms in which in which the loop momenta $q$ and $-q$
are inserted on the $p$ leg, as in a self-energy insertion, leading to terms which diverge like $1/p^2$.
Normally one would amputate this a-la LSZ, defining
\be
 S(p_i) = (\Pi_i Z_i^{1/2} p_i^2) \tilde{A}(p_i), \nonumber
\ee
where the $Z_i$ are appropriate wavefunction renormalization factors, $S(p_i)$ is a S-matrix element
and $\tilde{A}(p_i)$ is the sum of graphs for a correlation function omitting self-energy subgraphs.
Now in the present context it is not clear how to ``amputate,'' especially if the data
one is given consists of the physical forward amplitudes summed over all graphs, as we assume.

%Actually, before discussing on-shell limits, one should first make sure that the forward amplitude
%exist off-shell.  This amounts to the cancellation of tadpole graphs,
%which we postpone to the next subsection.

%In a supersymmetric theory, in practice this is a condition on the absence of gravitational
%anomalies
% external states which  is to say that no tadpoles arise.
% In practice this is a condition on the absence of gravitational anomalies,
% $\Tr t^a=0$, see below.)

The singular terms arise from well-separated self-energy subgraphs:
thus the analysis boils down to a 4-point function.
Including only the neighboring propagators, the forward amplitude of a gauge boson
against our gaugino reads
\be
 g^2C_A \frac{\pslash}{p^2} \times \frac12 \sum_\lambda \left[
   \es_\lambda \frac{\pslash {+} \qslash}{(p{+}q)^2} \es_\lambda^* 
 + \es_\lambda^* \frac{\pslash {-} \qslash}{(p{-}q)^2} \es_\lambda \right]\frac{\pslash}{p^2}
 \nonumber
\ee
where $(\epsilon_\lambda,q)$ describe the states of the (forward) gauge boson.

%Notice that in this case the flipped-helicity term comes term with a different color ordering.
%The reason we have distinguished the polarizations for $q$ and $-q$ will become clear below.

Upon summing over gauge boson polarizations
one finds the divergent contribution to the forward amplitude
\be
 -2 g^2C_A \times \frac{\pslash}{p^4}
\label{div1}
\ee

Such a divergence is interpreted as a gaugino mass being generated.  This may seem puzzling,
given that this should be forbidden by gauge invariance and Lorentz invariance.
However, one cannot invoke Lorentz invariance at this stage, since the $q$ integral is not performed yet!
In fact, there is no way the forward amplitude could have vanished at the integrand level:
in the presence of a thermal gas of gauge bosons, the gaugino would rightfully acquire a (non-Lorentz-invariant)
thermal mass, and it is precisely the job of the forward amplitude to generate it \cite{elmfors}.
What happens in vacuum is simply that \nr{div1} integrates to zero
by virtue of Lorentz invariance.  (The integrand is a total derivative,
$\int_q \delta(q^2)\propto \int_q \frac{\partial}{\partial q^\mu} q^\mu \delta(q^2)=0$, and, as such, it should
vanish in any reasonable regularization.)

Gauge invariance does not suffice to obtain cancellations at the integrand level.
Supersymmetry, however, provides such cancellations.
In our case, the superpartner of the gauge boson produces
the forward amplitude of a gaugino described by $(\lambda_q,q)$
\be
 g^2C_A \frac{\pslash}{p^2} \left[
   \gamma^\mu \lambda_q  \frac{1}{(p{+}q)^2} \tilde\lambda_q \gamma_\mu
  -\gamma^\mu \tilde\lambda_q^*  \frac{1}{(p{-}q)^2}
      \lambda_q^* \gamma_\mu \right]\frac{\pslash}{p^2}. \nonumber
\ee
This diverges like
% Attractive interaction in phi^3 would be another example.
% In a gravitational theory you will get a negative amplitude from boson-boson
% scattering by graviton exchange. That should be swamped out by positive contributions?
\be
 2g^2C_A \times \frac{\pslash}{p^4}. \nonumber
\ee
This exactly cancels the bosonic divergence, so that the sum possesses the milder $1/p^2$
behavior required for the LSZ amputation procedure.

It can be seen that these cancellations are general features of supersymmetric theories.
The point is that divergences are proportional to $1/p^4$ times
physical, on-shell gauge-invariant four-point scattering amplitudes.  Such forward four-point amplitudes
are just $c$-numbers, since all non-vanishing Mandelstam invariants are equal.
In supersymmetric theories one has to sum over the
polarizations of two of the four particles, $q$ and $-q$.  The sum will vanish provided the other two states,
$p$ and $-p$, form a BPS object.
For a massless particle in $\mathcal{N}=1$ this is the case
(the two supersymmetry transformations
$\lambda_p^\alpha Q_\alpha$ and $\tilde{\lambda}_p^{\dot\alpha}\tilde{Q}_{\dot\alpha}$ annihilate both states),
and for any particle
in $\mathcal{N}=2$ supersymmetry this is also the case.
Equality of asymptotic thermal masses in a finite-temperature context for
different particle species occurs by a similar mechanism \cite{mysusypaper}.
In all cases here, the mass shifts cancel.

After the leading divergence is so canceled, it is important that the subleading term be a clean $1/p^2$,
as required for LSZ reduction.  This is easily verified to be the case quite generally, as a consequence
of the Lorentz-invariance of the equations of motion of the theory -- the self-energy sandwiched between propagators
(which provide $\pslash$ polarization tensors) has to vanish when $p^2=0$, and after $1/p^4$ is canceled
we find that one is always left with a clean $1/p^2$.

In summary, in sufficiently supersymmetric theories the following limit exists
\be
 \Pi_i Z_i^{1/2} S(p_1,\ldots,p_n,q,-q) = \lim_{p_i^2\to m_i^2}
   \Pi_i [p_i^2 {+}m_i^2] \sum_h (-)^{2h}R_{n{+}2}(p_1,\ldots,p_n,q^h,-q^h).
\ee
The wavefunction factors $Z_i$ can be defined through a similar limit
\be
 Z_i^{1/2} \approx 1 + \frac12 \lim_{p_i^2\to m_i^2} [p_i^2{+}m_i^2] \sum_h (-)^{2h} G_4(p_i,-p_i,q^h,-q^h)
\ee
where $G_4$ is a four-point correlation function.  The combination of the two objects
produces correct S-matrix elements $S$ (with retarded boundary condition).

The gauge dependence in $Z_i$ and in the on-shell limit of $R$ cancel against each other,
making $S$ gauge-invariant as usual.  In fact we have checked, using Ward identities,
that this cancellation does takes place already at the level of the forward amplitude -- longitudinal gauge bosons
and ghosts cancel and need not be included in the helicity sum.

\subsubsection{Forward limits}

The following method might be a convenient alternative
way to evaluate the forward amplitudes.  The reader should be aware, however,
that we have only tested it in $\mathcal{N}=4$.

This method takes the external momenta to be on-shell from the start, which makes gauge invariance fully manifest,
but takes the amplitude to be slightly non-forward
\be
 \lim_{\epsilon\to 0} \sum_h (-)^{2h} R_{n{+}2}^\textrm{tree}(p_1,\ldots,p_n,q^h,(-q{+}\epsilon \mu)^{-h})
\ee
where $q{\cdot}k=0$ and now $R_{n{+}2}$ is already amputated.
It is easy to deform the other $p_i$ so as to conserve total momentum
and simultaneously preserve mass-shell conditions $p_i^2=-m_i^2$, as in section \ref{sec:pinch};
because mass shift divergences now cancel any such choice is equivalent.
\emph{A-priori}, now, this limit may depend on a arbitrary relative phase between the polarization
tensors of $q$ and $(-q{+}\epsilon k)$.  This will disappear if
complex-conjugate polarizations are used for CPT-conjugate states,
as is natural: then the ambiguity will be in the \emph{cosine} of a phase and drop out.
Provided the $\mu$-dependence drops out, which we have explicitly verified in $\mathcal{N}=4$,
this gives a reasonable candidate for the correct answer.

\subsection{Good ultraviolet behavior and factorization from unitarity cuts}

Consider deforming a forward amplitude in a gauge theory (we will consider only non-gravitational theories,
for simplicity) by taking
$q_1\to q_1+z\mu$ where $q_1{\cdot}\mu=\mu^2=0$.
\be
 R_{n{+}2}(z)\equiv \sum_h (-)^{2h} R_{n{+}2}(p_1,\ldots,p_n; (q+z\mu)^h, (-q-z\mu)^h).  \label{defRz}
\ee
Here again $R_{n{+}2}$ is already amputated and all momenta are on-shell.
This defines a rational function of complex $z$ whose poles are
due to internal propagators becoming on-shell.
This is similar to a Britto-Cachazo-Feng-Witten (BCFW) deformation \cite{BCFW}, but somewhat
simpler because really only one momentum is shifted.
As in the BCFW case, $R_{n{+}2}(z)$ is completely determined by its poles provided it vanishes
at infinity, which leads us to consider its behavior at infinity.

Direct inspection of individual Feynman graphs reveals
that $R_{n{+}2}(z)$ grows at most as $z$ at large $z$.  This
is the familiar eikonal limit.  (Here polarization tensors can be taken to scale like $\sim 1$
in this limit.) 
However, since the eikonal contribution is spin-independent it cancels among superpartners,
making $R_{n{+}2}(z)$ is at most of order one.

To find the true behavior of $R_{n{+}2}(z)$ a more refined analysis is needed.
Supersymmetry cancellations can be enforced from the start
by using $\mathcal{N}=1$ supersymmetry to trade
the sum over superpartners to a sum over supersymmetry-transformed
backgrounds,
\be
 R_{n{+}2}(z) = -\sum_{h~\textrm{bosonic}}
  \frac{\lambda_\alpha Q^\alpha}{\lambda_\alpha \lambda_q^\alpha}
  R_{n{+}2}(p_1',\ldots,p_n'; (q+z\mu)^h, (-q-z\mu)^{h-\frac12}).  \nonumber
\ee
where $Q$ acts only on the $p$ states;
$\lambda$ is an arbitrary reference spinor such that the denominator is nonzero,
where $\lambda_q(z)$ is defined by $\qslash(z) \lambda_q(z)=0$.
This is now a ``forward'' amplitude in which a boson turns into a fermion.
The leading diagrams have one gaugino exchange and a certain number
of gauge boson exchanges.
($F$-term interactions are not relevant because they couple, for instance,
a holomorphic scalar $\phi$ to holomorphic fermions $\psi$, while 
it is the complex conjugate fermion $\tilde\psi$ that appears in
the forward amplitude.)
In fact, gauge bosons can also be neglected since the light-cone gauge $q{\cdot}A=0$
is non-singular for the relevant diagrams.  Hence the large $z$ limit
is contained in a single tadpole graph involving the exchange of a zero-momentum gaugino,
and proportional to $\Tr t^a$.

Such a tadpole graph, if nonvanishing, would of course be ill-defined.
In fact, when such tadpoles graphs with zero-momentum propagators exist,
the forward amplitudes themselves are ambiguous to begin with, and what we
are seeing at large $z$ is just this ambiguity.  This possibility was ignored in
the last subsection;
in supersymmetric theories it can happen only if there is a gravitational
anomaly ($\Tr t^a\neq 0$).
Ignoring this possibility and assuming the absence of a gravitational anomaly,
then forward amplitudes exist and
the above argument shows that they vanish at infinity.

This implies that forward amplitudes can be constructed from their poles:
\ba
\hspace{-0.8cm} R_{n{+}2}(q) &=& \sum_I \int d\eta_1 d\eta_2 ~
   A_L(\{q+z(I)\mu,\eta_1\},\{p_I-(q+z(I)\mu),\eta_2\},\eta_{i_L})
\nl &\times&
   \frac{1}{(p_I{+}q)^2}
    A_R(\{-(q+z(I)\mu),-\eta_1\},\{-p_I+(q+z(I)\mu),-\eta_2\},\eta_{i_R}).
 \label{BCFW}
\ea
Each term is associated to a 2-particle factorization channel $I$ with total momentum $p_I$,
that is a 1-loop unitarity cut, with
$z(I)=-\frac{2p_I{\cdot}q + p_I^2}{2p_I{\cdot}\mu}$
corresponding to the point where $(p_I+q+z\mu)^2=0$; the $\eta$ variables are Grassmann variables
to do the sum over internal states.  This nicely expresses forward
amplitudes (in either $D$ or 4 dimensions), and thus loops, in terms of the integrands of
unitarity cuts (in $D$ or 4 dimensions, respectively).  No (bosonic) integration over these unitarity cuts
is needed to find the integrands.

Although we did not make use this fact here to obtain \nr{BCFW},
it may be useful to note that in supersymmetric theories only $4$-dimensional integrands are needed \cite{bddk}.

Two classes of theories have been left apart in the preceding two subsections:
theories with a gravitational anomaly, and, more interestingly, $\mathcal{N}=1$ theories
with massive particles.  On the other hand, \nr{BCFW} would appear to provide a
\emph{definition} of forward amplitudes in these two cases.  It would be interesting to further investigate
this possibility.

\subsection{$\mathcal{N}=4$ SYM examples}
\label{sec:n4sym}

It is illuminating to illustrate these findings in the most supersymmetric field theory,
$\mathcal{N}=4$ SYM.  For the present work we shall content ourselves with two simple 1-loop
examples.

\subsubsection{4-point function}

Consider the 4-point amplitude $\hat{R}_4(1^+,2^+,3^-,4^-)$ in planar $\mathcal{N}=4$ SYM.
Four different color structures contribute in the planar limit.  Let us focus on the
one involving $\hat{R}_6(-q,q,1^+,2^+,3^-,4^-)$.

The factorization \nr{BCFW} of this amplitude yields 3 terms.
It would be possible to evaluate them, but, actually, in $\mathcal{N}=4$ SYM, a much simpler
factorization exists.  This factorization involves taking $\mu= q$, e.g. deforming $q\to zq$.
In general this would lead to double poles at $z=0$
(corresponding to the familiar soft divergence $\sim 1/q{\cdot}p_1q{\cdot}p_4$),
which is not very convenient.  However,
in $\mathcal{N}=4$ SYM, a simple argument given in Appendix \ref{app:decay}
shows that the amplitude decay like $1/z^3$ at large $z$.  This is in keeping
with the finiteness of the theory.

The ``bonus'' obtained from this good decay is that one can 
factor the function $z^2 \hat{R}_6(zq)$ instead.
This function has no pole at the origin, leaving only a single channel
\be
 \hat{R}_6(q) = \frac{z_0^2}{(q{+}p_1{+}p_2)^2}
 \sum_h (-)^{2h} R_\textrm{left}(z_0q) R_\textrm{right}(z_0q),  \label{Rq}
\ee
where $z_0$ solves $(z_0 q{+}p_1{+}p_2)^2=0$ and both factors are MHV
four-point functions.
Recalling the MHV four-point function
\be
 \hat{R}_4(1^+,2^+,3^-,4^-) = \frac{\b{23}^4}{\b{12}\b{23}\b{34}\b{41}},
\ee
(and using its simple supersymmetric counterparts) we have
\be
 \sum_h (-)^{2h} R_\textrm{left}(z_0q) R_\textrm{right}(z_0q) =
\frac{ (\b{2q}\b{35} - \b{25}\b{3q})^4 }
{
 (\b{12}\b{25}\b{5q}\b{q1})
 (\b{34}\b{4q}\b{q5}\b{53})
 }   \label{factor4pt}
\ee
where $p_5$ is the on-shell momentum $z_0q{+}p_1{+}p_2$.
Here $\b{ij}=\epsilon_{\alpha\beta}\lambda_i^\alpha \lambda_j^\beta$
is the ordinary spinor product with $\epsilon_{12}=1$ the antisymmetric symbol.
(Null four-momenta are expanded as $p_i^{\dot{\alpha}\alpha}=\tilde{\lambda}_i^{\dot{\alpha}}\lambda_i^\alpha$.)
Using the Schouten identity, the numerator becomes simply
$\b{23}^4\b{q5}^4$.  Finally, using that
$\langle 5| \propto [q(p_1{+}p_2)|$, we obtain
\be
 \hat{R}_6(q) = \frac{\b{23}^4}{\b{12}\b{23}\b{34}\b{41}} \times 
\frac{-st}{(q{+}p_1)^2 (q{+}p_1{+}p_2)^2 (q{-}p_4)^2}.
\ee
The first factor is the tree amplitude.
The three other color structures are computed similarly.
Summing them, we ``discover'' that
\be
 \hat{R}_4^\textrm{1-loop} = \hat{R}_4^\textrm{tree} \times -g^2N st
\int_q \pi\delta(q^2) \left[
 \frac{1}{(q{+}p_1)^2(q{+}p_1{+}p_2)^2)(q{-}p_4)^2}
+ \textrm{cyclic}\right] !
\nonumber %\label{resSYM}
\ee
The sum is over the four cyclic permutations of $(1234)$.

In this expression we recognize the scalar box in disguise:
\be
 \hat{R}_4^\textrm{1-loop} = \hat{R}_4^\textrm{tree} \times -g^2N st
\int_q \frac{-i}{q^2 (q{+}p_1)^2 (q{+}p_1{+}p_2)^2 (q{-}p_4)^2}.
 \label{resSYM1}
\ee
That way, the present formalism quite simply leads to the well-known answer \cite{green,bddk}.
The integral will not be evaluated here, as we keep our focus on the derivation of the integrand.
Note that \nr{resSYM1} was obtained here without any a-priori assumption about its form.
We did not even assume that it could be expanded in terms of boxes.

A clarification is in order.  While we have evaluated the scattering amplitudes
in $4$ dimensions, the integral is $D$-dimensional.
As explained in \cite{bddk}, neglecting the dependence of the integrands
on the $(-2\epsilon)$ extra dimensions is justified whenever this dependence does not
overlap with ultraviolet divergences, which is totally expected in $\mathcal{N}=4$ SYM.

%\begin{figure}
%\begin{center}
%\includegraphics[width=14cm]{cont.eps}
%\end{center}
%\caption{The three terms in the BCFW-like factorization of $R(zq)$ along
%$q$ and $-q$.
%Only (b) contributes to the factorization of $z^2R(zq)$.
%}
%\%label{fig:SYM}
%\end{figure}

\def\NN{\mathcal{N}}
\subsubsection{5-point function}
\label{sec:5pt}

Let us present a slightly less trivial example, the 5-point MHV amplitude.
The integrand may always be written as
\be
 \frac{\hat{R}_5^{\textrm{1-loop}}}{\hat{R}_5^{\textrm{tree}}}
 = -g^2 \int_q \frac{\NN}{q^2(q{+}p_1)^2(q{+}p_1{+}p_2)^2(q{-}p_5{-}p_4)^2(q{-}p_5)^2}
 \label{def5pt}
\ee
where $\NN$ is a polynomial in the loop momentum $q$.
We will now determine $\NN$ from the forward amplitudes.
$\NN$ does not depend on external polarizations, as
MHV amplitudes with different polarizations are related to each other via supersymmetry.

Consider the forward amplitude in the color channel between particles 5 and 1,
summed over a $\mathcal{N}=4$ multiplet of states.
This can be obtained from a 2-term expression similar to \nr{factor4pt} or by any other means,
giving
\be
  \hat{R}_5^{\textrm{tree}} \times \left[ \frac{(q{+}p_{12})^2 s_{54}s_{51}
   +(q{-}p_5)^2 s_{21}s_{23}
 - \Tr_L \left[ qp_1p_2p_3p_4p_5 \right]}
 {(q{+}p_1)^2(q{+}p_{12})^2(q{-}p_{45})^2(q{-}p_5)^2}\right].
\ee
Here $\Tr_L[q p_1p_2p_3p_4p_5] \equiv \l q1\r [12]\l 23\r[34]\l45\r[5q]$
(in our convention in which the MHV tree amplitude depends only on angle brackets).
%The numerator may be written in the more symmetrical-looking form
%&=&
%   \frac12 (q{+}p_1)^2 s_{43}s_{45}
%   +\frac12 (q{+}p_{12})^2 s_{54}s_{51}
%   +\frac12 (q{-}p_{45})^2 s_{15}s_{12}
%   +\frac12 (q{-}p_5)^2 s_{21}s_{23}
%\nl &&
%   -\frac12 \Tr[\gamma_5 q12345]%
%\ea
Now this expression is defined only when $q^2=0$ but it must agree with the residue of \nr{def5pt}
at $q^2=0$ for any such $q$.
This fixes $\NN$ up to an ambiguity proportional to $q^2$:
\be
 \NN = (q{+}p_{12})^2 s_{54}s_{51}
   +(q{-}p_5)^2 s_{21}s_{23}
 - \Tr_L \left[ qp_1p_2p_3p_4p_5 \right] + q^2 C_1. \label{5pt1}
\ee
The ambiguity $C_1$ is $q$-independent because
the single cuts in other channels must decay like $1/q^3$.
The same computation in the channel between particles 1 and 2 shows that
\be
 \NN =
(q{-}p_{45})^2 s_{15}s_{12}
   +q^2 s_{32}s_{34}
 - \Tr_L \left[ qp_2p_3p_4p_5p_1 \right] + (q{+}p_1)^2 C_2 \label{5pt2}
\ee
where $C_2$ is another $q$-independent coefficient.
Equations \nr{5pt1} and \nr{5pt2} must both hold for any value of $q$,
and together they fully determine $\NN$.
Indeed, \nr{5pt2} uniquely determines $C_1$ and, \emph{vice-versa}, \nr{5pt1} uniquely determines $C_2$.
After a short bit of algebra one finds
\ba
\NN = (q{+}p_{12})^2 s_{54}s_{51}
   +(q{-}p_5)^2 s_{21}s_{23}
   + \frac12 q^2 s_{32}s_{34}
 - \Tr_L \left[ q p_1p_2p_3p_4p_5 \right]
 - \frac12 q^2 \Tr\left[\gamma_5 p_2p_3p_4p_5\right].  \nonumber
\ea
This together with \nr{def5pt} constitute our final expression for the 5-point MHV 1-loop integrand.
By construction, it is rigorously equal to the
gauge-invariant sum of order $\sim 100$ 1-loop Feynman diagrams, modulo purely algebraic manipulations.

It is possible to rewrite this result in somewhat more symmetrical form,
\ba
   \NN &=&
    \frac12 q^2 s_{32}s_{34}
   +\frac12 (q{+}p_1)^2 s_{43}s_{45}
   +\frac12 (q{+}p_{12})^2 s_{54}s_{51}
   +\frac12 (q{-}p_{45})^2 s_{15}s_{12}
   +\frac12 (q{-}p_5)^2 s_{21}s_{23}
 \nl &&
   - \frac12 \Tr[\gamma_5 q p_1p_2p_3p_4p_5]
   - \frac12 q^2\Tr[ \gamma_5 p_2p_3p_4p_5].
\ea
The first line gives a sum of five 1-mass boxes
and agrees precisely with the form of the integrand obtained long ago
in \cite{bddk}, using unitarity-based methods.

The second, parity-odd, line is not manifestly
cyclically invariant; it nevertheless is.
Its overall sign would be opposite for a $\overline{\mbox{MHV}}$ amplitude.
Using Passarino-Veltman reduction, however, it is possible to show that it integrates to $\OO(\epsilon)$,
which is why this contribution is usually not quoted.
In that sense, integrands (and their single-cuts)
contain more information than any of their integrals.

We have similarly checked that the formalism works for general $n$-point MHV amplitudes.\footnote{
 In that case, in addition to the familiar sum over 2-mass easy boxes, the formalism also produces
 the parity-odd pentagons which were discussed in \cite{cachazo}.
 Although these pentagons integrate to zero, they cannot be dropped through any sequence of algebraic manipulations.
}

It is striking that forward amplitudes in only two channels ``know'' about the
amplitude in all other channels.  The magic, evidently, lies
in the fast decay of forward amplitudes in $\mathcal{N}=4$ SYM.
Using this decay $\sim 1/q^3$, actually, it seems very likely that one could iterate the thinking underlying \nr{BCFW}
to determine individual the BCFW terms in terms of triple-cuts, which would then be determined by quadruple-cuts,
at which point one would run out of decay properties.
Thus forward amplitudes in $\mathcal{N}=4$ should be determined by the leading singularities of the theory.\footnote{
 More precisely, the data here would be the leading singularities of the integrand
 on all possible contours, as opposed to only the parity-even combinations which are sometimes considered.
}
For this reason we expect that, in general, forward amplitudes in only $(n{-}3)$ channels should suffice to determine
$n$-point 1-loop $\mathcal{N}=4$ integrands.

\section{Summary and outlook}

Exploiting the physical principles of causality and locality, we have obtained
relations between loop amplitudes and the forward limit
of tree amplitudes, which hold for correlation functions in any quantum field theory.
At 1-loop the relations are known as Feynman's tree theorem, \nr{1loop}.
Use of causal boundary conditions has allotted us significant simplifications to the formulas
and discussion, while sacrificing none of the physics.
A simple analytic continuation relates retarded amplitudes and time-ordered amplitudes,
but retarded amplitudes could also be useful on their own for computing inclusive observables \cite{gelisvenugopalan}.

The generalization of the tree theorem to higher-loop is nontrivial.
Only for planar gauge theories does one gets physical tree amplitudes.
In that case our results allow the interpretation of the
multi-loop vacuum as a collection of on-shell particles
with completely classical behavior.  In the infinite past (where interactions are turned off),
these particles exist in an intricately
entangled but universal color state.  We have computed this state up to the 3-loop approximation
but we attempted no 4-loop computation.
Already at 2-loop, this color state involves interesting factors of $\frac13$ as given
in \nr{R2loopcolor}.

In non-planar theories a similar picture holds at 1-loop but not to higher-loop.
Our discussion of non-planar theories is still rather preliminary and we leave it
in a somewhat puzzling state.
On the one hand we presented a strong argument in section \ref{sec:guesswork},
that integral representations in Minkowski space involving
physical forward amplitudes do not exist then.  We believe that this argument
is insensitive to the boundary conditions used, retarded or time-ordered.
On the other hand, we proposed that loop integrals are nevertheless implicitly
determined by physical forward amplitudes.
This conjecture, if true, would allow for a purely algebraic
determination of higher-loop integrands starting from physical forward amplitudes.
That would be complimentory to unitarity-based approaches, in that one also starts from
an Ansatz for the loop integrand and then fixes the coefficients.
The proper physical interpretation of the conjecture, if true, is still unclear.
Our support for it comes from a computation in complexified Minkowski space-time
that we presented in section \ref{sec:uniqueness}.

The conjecture is also supported by the fact that it would be implied,
if unitarity indeed suffices to determine loops.
Since unitarity-based techniques have been successfully applied to even 4-loop
in a non-planar context ($\mathcal{N}=8$ supergravity) \cite{bernN8},
this is actually fairly strong support.
It is important to stress that these considerations are rather formal.  In the end,
the only real test of the conjecture will be to work out explicit non-planar higher-loop examples.

We have deliberately left out of the discussion the computation of forward amplitudes
for S-matrix elements.  For off-shell correlation functions, to which the formalism applies,
we believe that everything is under control.
But for S-matrix elements, probably the most interesting observables,
LSZ reduction at the integrand level must still be thought through carefully.
For supersymmetric theories we did work it out fully, in section \ref{sec:susy},
and this gives us hope that technical difficulties in the general case
will be solved in the future.

Nowhere in this paper did we use that the theory was in a Lorentz-invariant state.
For this reasons, all our results extend to finite-temperature contexts.
Since we used the Schwinger-Keldysh formalism this is actually
quite straightforward -- formally one just needs to add physical particles to
the vacuum ones, specifically $\pm\pi\delta(q^2)\to 2\pi\delta(q^2)[\pm\frac12 +n(q)]$,
where $n(q)=1/(e^{q^0/T}\mp 1)$ are particle distributions as usual,
in \nr{1loop} and in the classical part of \nr{R2loop} at 2-loop.
In that case our formulas are conceptually very close, albeit distinct, from
other formulas such as
the celebrated Dashen-Ma-Bernstein formula for the thermodynamic potential \cite{DMB} and also
the correlation functions formulas in \cite{jeonellis,blaizot04}.
These formulas involve vacuum loop amplitude, while the present ones involve vacuum tree amplitudes.

Finally, it is worth stressing that for sufficiently supersymmetric planar gauge theories,
no technical subtleties arise and the formulas in this paper directly give
loop-level S-matrix elements, at least up to 3-loop, as integrals over trees.
These integrands manifest all symmetries of the tree amplitudes.
Certainly, it will be interesting to further study these integrands.

\emph{Note:} While this paper was being completed, works \cite{catani2} appeared in which
$i\epsilon$ prescriptions are discussed in the context of cutting open higher-loop graphs.
In the present work we avoided such questions by using retarded boundary conditions,
which make the $i\epsilon$'s clear and unambiguous,
using analytic continuation in the end to convert to time-ordered conditions.

\section*{Acknowledgments}

The author is grateful to Nima Arkani-Hamed, Freddy Cachazo and Sangyong Jeon for
useful discussions at various stages of this project.
This work was funded in part by NSF grant PHY-0503584.

\begin{appendix}

\section{Schwinger-Keldysh Feynman rules}
\label{app:rules}

\begin{figure}
\begin{center}
\includegraphics[width=10cm]{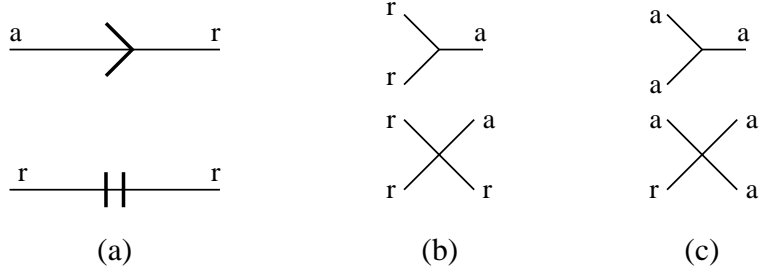}
\end{center}
\caption{
 Schwinger-Keldysh Feynman rules. (a) retarded and cut propagators.
 (b) Ordinary vertices, that have one $a$ index.
 (c) $\hbar^2$-suppressed vertices, that have three $a$ indices.
}
\label{fig:rules}
\end{figure}

The Schwinger-Keldysh Feynman rules are summarized in Fig.~\ref{fig:rules}.
The propagator is the $2\times 2$ matrix
\be
  \left(\begin{array}{ll}
  G^{rr} & G^{ra} \\ G^{ar} & G^{aa}\end{array}\right)
 = \left(\begin{array}{ll}
  \frac12 (G^>+G^<) & \GR \\ \GA & 0 \end{array} \right)
 \simeq \left(\begin{array}{ll}
  \pi\delta(p^2+m^2) & \frac{-1}{p^2-i\epsilon p^0} \\ \frac{-1}{p^2+i\epsilon p^0}
  & 0 \end{array} \right).
\nonumber
\ee
We find convenient to draw retarded propagators, which form the two off-diagonal elements,
with an arrow on them representing the time flow.  We draw a cut on the ($\sim \hbar$)
symmetrical propagator $G^{rr}$, which is the only diagonal element.
Interaction vertices with one outgoing arrow (such as $rra$ or $rrra$)
are the usual ones, and vertices with tree outgoing arrows (such as $aaa$ or $raaa$)
are down by $\hbar^2/4$.

\section{The scalar triangle in $D$ dimensions}
\label{app:triangle}

In this appendix we consider the triangle scalar integral in $D$ dimensions
with 3 generic external momenta.  We want to check that the single-cut formalism
reproduces it correctly.  It is expressed in the form
\be
 \int_q \pi\delta(q^2)\left[
 \frac{1}{(2q{\cdot}p_1 {+} p_1^2)
          (-2q{\cdot}p_2 {+} p_2^2)}
{+} \frac{1}{(2q{\cdot}p_2 {+} p_2^2)
          (-2q{\cdot}p_3 {+} p_3^2)}
{+} \frac{1}{(2q{\cdot}p_3 {+} p_3^2)
          (-2q{\cdot}p_1 {+} p_1^2)}\right].
% \nonumber 
\label{triangleA2}
\ee

We will do these integrals using Feynman parameters.
The basic integral is
%\be
% \frac{1}{(2q{\cdot}p_1 {+} p_1^2)
%          (-2q{\cdot}p_2 {+} p_2^2)} =
%\int_0^\infty da \frac{1}{ (2q{\cdot}(p_1-ap_2) {+} p_1^2 {+} ap_2^2)^2 },
% \label{Feynmanparm}
%\ee
% XXX give phase space form.
\be
 \int_q \frac{\pi\delta(q^2)}{[2q{\cdot}p {+} k^2]^n} =
%   \frac{-\Gamma(2\epsilon)\Gamma(2 + 2\epsilon)\Gamma(\frac12)}
% {2^{1-2\epsilon} (4\pi)^{\frac{D}2} \Gamma(\frac32-\epsilon}}
  -\frac{c_\Gamma}{2\epsilon}
 \times \frac{1}{ (p^2)^{1-\epsilon} (k^2)^{n{-}2{+}2\epsilon}}
\times   \frac{\Gamma(n{-}2{+}2\epsilon)}{\Gamma(2\epsilon)\Gamma(n)}, \nonumber
\ee
where $c_\Gamma= \frac{\Gamma(1+\epsilon)\Gamma(1-\epsilon)^2}{(4\pi)^{2-\epsilon} \Gamma(1-2\epsilon)}$.
The branch of this function will be discussed momentarily.

Proceeding formally for the moment, the first term in the bracket becomes
\ba
 \int_q  \frac{\pi\delta(q^2)}{(2q{\cdot}p_1 {+} p_1^2)
               (-2q{\cdot}p_2 {+} p_2^2)}
&=& \int_q \int_0^1 dy \frac{\pi\delta(q^2)}{[ 2q{\cdot} (yp_1{-}(1{-}y)p_2) + yp_1^2+(1{-}y)p_2^2]^2}
\nl
&=& -\frac{c_\Gamma}{2\epsilon} \frac{1}{(p_1^2p_2^2)^\epsilon}
   \int_0^1 \frac{dx}{(xp_1^2 + (1{-}x)p_2^2 -x(1{-}x) p_3^2)^{1-\epsilon}}
\nonumber
\ea
where on the second line we have changed variables to $y=(1{-}x)p_1^2/( (1{-}x)p_1^2+xp_2^2)$.
In Euclidean kinematics ($p_i^2>0$ for $i=1,2,3$), the denominator is real
and positive everywhere along the naive contour which has $x$ real.
The $x$ integral along that contour can
be evaluated explicitly in terms of Gauss' hypergeometric function:
\ba
 -\frac{c_\Gamma}{2\epsilon^2} \frac{i}{(p_1^2p_2^2p_3^2)^\epsilon
     \Delta^{\frac12 - \epsilon}} &&
\nl
&& \hspace{-5cm} \times
 \left[
   \left( \frac{1{+}i\delta_1}{1{-}i\delta_1}\right)^\epsilon
       {}_2F_1(\epsilon,2\epsilon,1{+}\epsilon, - \frac{1{+}i\delta_1}{1{-}i\delta_1})
%\right.} \nl
% && \left.
  -\left( \frac{1{-}i\delta_2}{1{+}i\delta_2}\right)^\epsilon
       {}_2F_1(\epsilon,2\epsilon,1{+}\epsilon, - \frac{1{-}i\delta_2}{1{+}i\delta_2})
 \right]. \label{hypergeom}
\ea
Here $\Delta$ is the is the so-called Gram determinant
\ba
 \Delta&=& 2p_1^2p_2^2 {+} 2p_1^2p_3^2 {+}2p_2^2p_3^2-p_1^4-p_2^4-p_3^4
   = 4 p_1^2p_2^2-4(p_1{\cdot}p_2)^2 \nonumber
\ea
and
$\delta_1= \frac{2p_2{\cdot}p_3}{\sqrt{\Delta}}$,
 with similar definitions for cyclic permutations.
Using hypergeometric identities
it is possible to show
that the bracket is symmetrical under interchange of $p_1$ and $p_2$.

The proper $x$ contour must now be discussed.
We consider, for simplicity, only the Euclidean regime where all external momenta are space-like $p_i^2>0$.
The integrand has two branch cuts, which begin at the two complex zeroes
of the denominator and reach out to infinity.
In the second and third term in \nr{triangleA2} the poles of the two factors are on the same
side of the real axis and so the ordinary real $x$ contour is correct.
In the first term, the boundary condition in the two factors are not consistent
which means that the real contour makes no sense.  Rather, the contour must go from 0 to infinity
and then from infinity to 1, so as to ensure that, as a function of $x$, the pole
in the loop integrand never cross the real frequency axis.
This contour differs from the real one by a discontinuity across one of the two branch cut
(either one produces the same discontinuity).
This produces the additional contribution
\be
 -\frac{c_\Gamma}{2\epsilon (p_1^2p_2^2)^\epsilon} \int_{x_0}^\infty
  \frac{dx}{ (p_3^2(x{-}x_0)(x{-}\overline{x}_0))^{1{-}\epsilon} }
%\nl
 =
   -\frac{c_\Gamma}{2\epsilon^2} \frac{i}{(p_1^2p_2^2p_3^2)^\epsilon
     \Delta^{\frac12 - \epsilon}}
 \frac{2\pi \Gamma(1-2\epsilon)}{\Gamma^2(1-\epsilon)}. \nonumber
\ee

Summing over all contributions, we therefore find
\be
 I_3 = \frac{c_\Gamma}{2\epsilon^2 (p_1^2p_2^2p_3^2)^\epsilon
     \Delta^{\frac12 - \epsilon}}
 [ f(\delta_1) + f(\delta_2) + f(\delta_3) + c] \nonumber
\ee
where
\be
 f(\delta_i) \equiv \frac{1}{i}
 \left[
   \left( \frac{1{+}i\delta_i}{1{-}i\delta_i}\right)^\epsilon
       {}_2F_1(\epsilon,2\epsilon,1{+}\epsilon, - \frac{1{+}i\delta_i}{1{-}i\delta_i})
%\right.} \nl
% && \left.
  -\left( \frac{1{-}i\delta_i}{1{+}i\delta_i}\right)^\epsilon
       {}_2F_1(\epsilon,2\epsilon,1{+}\epsilon, - \frac{1{-}i\delta_i}{1{+}i\delta_i})
 \right] \nonumber
\ee
and
\be
 c= -2\pi\epsilon \frac{\Gamma(1-2\epsilon)}{\Gamma^2(1-\epsilon)}. \nonumber
\ee
The last term is the branch cut contribution.
This agrees precisely with the form given in \cite{bdkpentagon}, Appendix V, confirming the validity
of the single-cut integrals.

More precisely, this is the result in Euclidean kinematics.  It can be converted
to Minkowski kinematics, with either retarded or time-ordered boundary conditions,
by means of the analytic continuations described in section \ref{sec:keldysh}.

For convenience we used the same notation as \cite{bdkpentagon}
 (up to a factor $p_1^2p_2^2p_3^2$ in our definition of $\Delta$).
This has a finite limit as $p_1^2$ and $p_2^2$ go to zero, which agrees with \nr{triangleres},
and also a finite limit as $\epsilon\to 0$ which can be expressed in terms of dilogarithms \cite{bdkpentagon}.

% The apparent sign mismatch is because BDK work in (+---) signature.

\section{Some terms in the 2-loop vacuum polarization numerator}
\label{app:ferm}

We list the contributions to $\PiR^{\mu\nu}(p)$ with two quark lines cut.
These are the forward amplitudes of two quarks against two currents
as depicted in Fig.~\ref{fig:2loop} (c).
We denote $q_1$ and $q_2$ the momenta of the two quarks.
\begin{align}
 - \int \pi\delta'(q_1^2)\pi\delta(q_2^2) \Tr&\left[
 q_1\gamma^\mu \frac{1}{q_1+p}\gamma^\nu q_1 \gamma^\alpha
 q_2\gamma_\alpha\frac{1}{(q_2{-}q_1)^2} \right.
\nl &\left. + ((p,\mu)\leftrightarrow (-p,\nu))\right]
\nl
 +\int \pi\delta(q_1^2)\pi\delta(q_2^2) \Tr&\left[
 q_1\gamma^\mu \frac{1}{q_1+p} \gamma^\alpha \frac{1}{q_2+p} \gamma^\nu
q_2 \gamma_\alpha \frac{1}{(q_1{-}q_2)^2} \right.
\nl
& \left.
 + q_1\gamma^\mu \frac{1}{q_1+p} \gamma^\alpha q_2
 \gamma_\alpha \frac{1}{q_1+p}\gamma^\nu \frac{1}{(q_1{+}p{-}q_2)^2}\right.
\nl
& \left.
 + ((p,\mu)\leftrightarrow (-p,\nu)) \right]
\nl
+\int \pi\delta(q_1^2)\pi\delta(q_2^2)\Tr&\left[
 q_1\gamma^\mu \frac{1}{q_1+p} \gamma^\alpha q_2
  \gamma^\nu \frac{1}{q_2-p} \gamma_\alpha \frac{1}{(q_1{+}p{-}q_2)^2} \right.
 \nl
  & \left.
 + q_1\gamma^\alpha \frac{1}{q_2-p} \gamma^\mu q_2
  \gamma_\alpha \frac{1}{q_1+p} \gamma^\nu \frac{1}{(q_1{+}p{-}q_2)^2} \right].
  \label{fermcut}
\end{align}
This expression is to symmetrized, with unit weight,
under the obvious symmetry $q_1\leftrightarrow q_2$, which we have
not manifested for conciseness of notation.
With this in mind, manifest transverseness of
the integrand with respect to $p_\mu$ is readily verified.

The sum of the first and fourth line,
which share the Dirac structure
$\Tr[\aslash{A} \gamma^\mu \aslash{B} \gamma_\mu \aslash{C}\gamma^\alpha \aslash{D}\gamma_\alpha]$
upon contracting with $\eta^{\mu\nu}$, is reproduced in \nr{long2}.
The remaining contributions have been included in the final
answer \nr{result}.

\section{A (simple) 2-loop scalar integral}
\label{app:2loop}

In this Appendix we consider the 2-loop scalar integral
\be
 I=\int_{q_1,q_2} \frac{(-i)^2}{q_1^2q_2^2} \frac{1}{(q_1{+}q_2)^2(q_1{+}p)^2} =
 d_\Gamma \frac{\Gamma(2\epsilon) \Gamma(1-\epsilon)\Gamma(1-2\epsilon)}{(4\pi)^{\frac{D}{2}}
 \Gamma(1+\epsilon)\Gamma(2-3\epsilon) } (p^2)^{-2\epsilon},
% \frac{\Gamma(\epsilon)\Gamma(2\epsilon)}{(4\pi)^D}
 % \frac{\Gamma(1-\epsilon)^3(p^2)^{-2\epsilon}}
 % {(1-2\epsilon)\Gamma(1+\epsilon)\Gamma(2-3\epsilon)}
  \label{Acgoal}
\ee
where
\be
 d_\Gamma = \frac{\Gamma(\epsilon)\Gamma(1-\epsilon)^2}
  {(4\pi)^{\frac{D}{2}}\Gamma(2-2\epsilon)}.
 \nonumber
\ee
We want to check that this result is correctly reproduced by \nr{R2loop},
which will provide a nontrivial
check on the factor $\frac13$ in that equation.

Applied to this integral, \nr{R2loop} produces the three terms
\begin{align}
 I=\int_{q_1,q_2} X(q_1,q_2)&\left[ \frac{1}{(q_1{+}q_2)^2(q_1{+}q_2{+}p)^2}
 +\frac{2}{(q_1{+}q_2)^2(q_1{+}p)^2} 
 \right.\nl&\left.+ \frac{2}{(q_1{+}q_2{+}p)^2(q_1{+}p)^2}\right],
 \label{mainAc}
\end{align}
where
\ba \nonumber
 X(q_1,q_2) &\equiv& 
   \pi^2\left[
  \frac43\delta^+(q_1^2)\delta^+(q_2^2)
  +\frac23\delta^+(q_1^2)\delta^-(q_2^2)\right.
\nl && \left. 
 +\frac23\delta^-(q_1^2)\delta^+(q_2^2)
+\frac43\delta^-(q_1^2)\delta^-(q_2^2)\right].
\nonumber
\ea
%We consider only $p$ timelike.
The $i\epsilon$ prescriptions in the denominators, not indicated,
may be recovered trivially by considering the time flow in the corresponding diagrams.

Our strategy is to first perform the integration over $q_2$ and then analyze the remaining
one over $q_1$.
The first term in the bracket depends only on the combination $(q_1{+}q_2)$ and for it we use that
\be
 \int_{q_1,q_2} X(q_1,q_2) (2\pi)^D\delta^D(q_1{+}q_2-q) =
%  \frac{2^{2\epsilon}\pi^\frac32 |q^2|^{-\epsilon}}
%   {3(4\pi)^{\frac{D}{2}}\Gamma(\frac{D{-}1}{2})}
%   \left[\theta(-q^2) + \frac{\theta(q^2)}{2\sin(\pi\frac{D-1}{2})}\right],
% \nl
  d_\Gamma \sin(\pi\epsilon) \left[ \frac23\theta(-q^2) 
 - \frac{\theta(q^2)}{3\cos(\pi\epsilon)}\right].
\nonumber % \label{dummyAc1}.
\ee
%The integral diverges linearly in the ultraviolet but
%it has no pole in $D=4$ dimensions.
For the remaining two terms, we use that
\be
 \int_{q_2}\frac{2X(q_1,q_2)}{(q_2{+}A)^2} = %-i\epsilon(q_2^0{+}A^0)} =
%%% Next thing is for X not 2X
% \pi\delta(q_1^2)  \frac{\Gamma(\epsilon)}{2(4\pi)^{\frac{d}{2}}}
% \frac{\Gamma(1-\epsilon)^2}{\Gamma(2-2\epsilon)}
% (A^2-i\epsilon A^0)^{-\epsilon}  \nl
% &&
% +i\pi[\delta^+(q_1^2)-\delta^-(q_1^2)] \frac{2^{2\epsilon}\pi(A^2-i\epsilon A^0)^{-\epsilon}}
% {24(4\pi)^{\frac{D-1}{2}}
% \Gamma(\frac{D-1}{2})\sin(\pi\frac{D-1}{2})}.
 d_\Gamma (A^2)^{-\epsilon}
\left[\frac{\pi\delta^+(q_1^2) {+} \pi\delta^-(q_1^2)}{2}
  -i\frac{\sin(\pi\epsilon)}{3\cos(\pi\epsilon)} 
 [\pi\delta^+(q_1^2)-\pi\delta^-(q_1^2)]\right].
\nonumber % \label{dummyAc2}
\ee
Actually, in the second term, $A^2=q_1^2=0$ and so this whole term vanishes
--- it contains a massless bubble with no scale in it.

Using those results, the first and third terms of \nr{mainAc} sum to
\begin{align}
I= d_\Gamma \int_{q}&
\left[ \frac{\sin(\pi\epsilon)\theta(-q^2)}{q^2|q^2|^\epsilon(q{+}p)^2} +
 \frac{\pi\delta(q^2)}{(q{+}p)^{2(1{+}\epsilon)}} \right]
\nl
-d_\Gamma \frac{\sin(\pi\epsilon)}{3\cos(\pi\epsilon)}\int_q
&
\left[ \frac{\theta(q^2) + \theta(-q^2)\cos(\pi\epsilon)}{q^2(q{+}p)^2|q^2|^{2\epsilon}}
+ i\frac{[\pi\delta^+(q^2)-\pi\delta^-(q^2)]}{(q{+}p)^{2(1+\epsilon)}}
\right].
\end{align}
We have chosen to group the terms in a specific way for reasons that will soon
become manifest.
The easiest way to evaluate the integral on the first line is to note that that it is equal
 to $\int_q (-i)/q^2(q{+}p)^{2(1{+}\epsilon)}$, as follows from applying contour integration
  on the $q^0$ integral in the latter expression.
In any case, it gives
\be
 I= d_\Gamma \frac{\Gamma(2\epsilon) \Gamma(1-\epsilon)\Gamma(1-2\epsilon)}{(4\pi)^{\frac{D}{2}}
 \Gamma(1+\epsilon)\Gamma(2-3\epsilon)} (p^2)^{-2\epsilon}. \label{resultA2}
\ee
The integral on the second line vanishes; the easiest way to see this is to note that,
in the frequency plane, it is equal to a horizontal contour integral
that can be deformed to infinity without crossing any singularity.
In any case, it vanishes.
Equation \nr{resultA2} agrees precisely with \nr{Acgoal}, which confirms in a
non-trivial way the factor $\frac13$ in \nr{R2loop}.

\section{Fast decay in $\mathcal{N}=4$ SYM}
\label{app:decay}

We show that forward amplitudes in $\mathcal{N}=4$ decay
like $1/q^3$ at large $q$.  The argument is a simple extension of
the one given in \cite{simplest}, to establish
the absence of one-loop bubbles in $\mathcal{N}=4$ and supergravity.

We begin with the representation \nr{BCFW} of the forward amplitude in terms of unitarity
cuts integrands,
\ba
 A_{n{+}2}(wq) &=& \sum_I \int d^4\eta_1 d^4\eta_2 ~
   A_L(\{wq+z(I)\mu,\eta_1\},\{p_I-(wq+z(I)\mu),\eta_2\},\eta_{i_L})
\nl &\times&
   \frac{1}{(p_I{+}wq)^2}
    A_R(\{-(wq+z(I)\mu),-\eta_1\},\{-p_I+(wq+z(I)\mu),-\eta_2\},\eta_{i_R}),
 \nonumber % \label{deform2}
\ea
where $z(I)=-\frac{2w p_I{\cdot}q + p_I^2}{2p_I{\cdot}\mu}$.

We study each term separately.  The main point is that
both $A_L$ and $A_R$ are themselves
BCFW-deformed amplitudes:
\be
 wq{+}z(I)\mu = w (q-\frac{p_I{\cdot}q}{p_I{\cdot}\mu} \mu) -\frac{p_I^2}{2p_{I}{\cdot}\mu} \mu.
 \nonumber
\ee
The can be interpreted as a BCFW deformation, the first parenthesis
giving the deformation vector.
Hence this places us precisely in the situation considered in \cite{simplest},
and it remains to estimate the fermion integral over $\eta_1$ and $\eta_2$,
or, equivalently, to sum over intermediate states.
This was done in \cite{simplest}, and we briefly review their argument.
Using a $z$-independent supersymmetry transformation,
$\eta_1$ and $\eta_2$ can be both set to zero in $A_L$ and $A_R$ at the cost of absorbing these variables
into a $w$-independent but $\eta$-dependent redefinition of the background.
$A_L$ times $A_R$ summed over a supermultiplet of states then become the
amplitudes of four $+$-helicity gluons against a combination of supersymmetry-transformed backgrounds.
Each such amplitude decays like $1/w$, and, including
the factor $1/w$ from the propagator, one finds $A(wq)\sim 1/w^3$ as claimed.
%This argument crucially needed all the supersymmetries of $\mathcal{N}=4$,
%in agreement with the fact that $\mathcal{N}=2$ theories are not finite.

\end{appendix}

\end{document}